\documentclass[sigconf]{acmart}

\usepackage{listings}
\usepackage{acmart-taps}

\AtBeginDocument{%
  \providecommand\BibTeX{{%
    \normalfont B\kern-0.5em{\scshape i\kern-0.25em b}\kern-0.8em\TeX}}}

\copyrightyear{2025}
\acmYear{2025}
\setcopyright{cc}
\setcctype{by}
\acmConference[UIST '25]{The 38th Annual ACM Symposium on User Interface Software and Technology}{September 28-October 1, 2025}{Busan, Republic of Korea}
\acmBooktitle{The 38th Annual ACM Symposium on User Interface Software and Technology (UIST '25), September 28-October 1, 2025, Busan, Republic of Korea}\acmDOI{10.1145/3746059.3747795}
\acmISBN{979-8-4007-2037-6/2025/09}

\makeatletter
\def\@ACM@copyright@check@cc{}
\makeatother

\definecolor{brown}{rgb}{0.59, 0.29, 0.0}
\definecolor{darkgray}{rgb}{0.59, 0.59, 0.59}
\definecolor{tablegray}{gray}{.9}

\definecolor{nicecolor}{rgb}{0, 0.58, 0.51}

\newcommand\amy[1]{\textit{\textcolor{nicecolor}{[Amy] #1}}}
\newcommand\karim[1]{\textit{\textcolor{purple}{[Karim] #1}}}

\newcommand\highlight[1]{\textcolor{red}}
\newcommand\revision[1]{#1}
\usepackage[utf8]{inputenc}
\usepackage{diagbox}
\usepackage{colortbl}

\usepackage{tabularx}
\usepackage{multirow}

\usepackage{xcolor,colortbl}
\usepackage{soul}

\usepackage{xspace}
\usepackage{enumitem}
\usepackage{mathtools}
\usepackage{commath}

\usepackage{amssymb}% http://ctan.org/pkg/amssymb
\usepackage{pifont}% http://ctan.org/pkg/pifont

\newcommand{\customtilde}{{\raise.17ex\hbox{$\scriptstyle\sim$}}}

\newcommand{\etal}{et~al.\xspace}
\newcommand{\eg}{\textit{e.\,g.},\xspace}
\newcommand{\ie}{\textit{i.\,e.},\xspace}

\usepackage{xparse}
\usepackage{mydefs}

\newcommand{\system}{TalkLess}

\definecolor{deletioncolor}{HTML}{eaecec}
\definecolor{insertioncolor}{HTML}{f6e5f1}
\definecolor{fillerwordcolor}{HTML}{dcefdc}
\definecolor{repetitioncolor}{HTML}{e1efc4}
\definecolor{emphasiscolor}{HTML}{faefb7}
\definecolor{clarificationcolor}{HTML}{f8ce88}
\definecolor{informationcolor}{HTML}{f9a77c}
\definecolor{deletedgray}{HTML}{7F7F7F}

\newcommand{\markerbox}[2]{\colorbox{#1}{\raisebox{0pt}[1.2ex][0.1ex]{#2}}}

\newcommand{\insertion}[1]{{\sethlcolor{insertioncolor}\hl{\textbf{#1}}}}
\newcommand{\fillword}[1]{{\sethlcolor{fillerwordcolor}\hl{#1}}}
\newcommand{\repeatword}[1]{{\sethlcolor{repetitioncolor}\hl{#1}}}
\newcommand{\emphword}[1]{{\sethlcolor{emphasiscolor}\hl{#1}}}
\newcommand{\clarify}[1]{{\sethlcolor{clarificationcolor}\hl{#1}}}
\newcommand{\info}[1]{{\sethlcolor{informationcolor}\hl{#1}}}

\definecolor{emphasislight}{HTML}{fcf6d2}  % Light (least emphasis)
\definecolor{emphasismedium}{HTML}{faefb7} % Medium
\definecolor{emphasisstrong}{HTML}{fce45a} % Strong (most emphasis)

\newcommand{\emphasisl}[1]{{\sethlcolor{emphasislight}\hl{#1}}}
\newcommand{\emphasism}[1]{{\sethlcolor{emphasismedium}\hl{#1}}}
\newcommand{\emphasiss}[1]{{\sethlcolor{emphasisstrong}\hl{#1}}}

\newcommand{\alphaval}[2]{{\small $p\,#1\,#2$}}
\begin{document}

\title{\system: Blending Extractive and Abstractive Summarization for Editing Speech to Preserve Content and Style}

\author{Karim Benharrak}
\orcid{0009-0002-3279-5664}
\email{karim@cs.utexas.edu}
\affiliation{%
  \institution{The University of Texas at Austin}
  \city{Austin}
  \state{TX}
  \country{USA}
}

\author{Puyuan Peng}
\orcid{}
\email{pyp@utexas.edu}
\affiliation{%
  \institution{The University of Texas at Austin}
  \city{Austin}
  \state{TX}
  \country{USA}
}

\author{Amy Pavel}
\orcid{0000-0002-3908-4366}
\email{amypavel@eecs.berkeley.edu}
\affiliation{%
  \institution{University of California, Berkeley}
  \city{Berkeley}
  \state{CA}
  \country{USA}
}

\renewcommand{\shortauthors}{Benharrak, et al.}
\renewcommand{\shorttitle}{\system: Blending Extractive and Abstractive Summarization for Editing Speech to \\Preserve Content and Style}

\begin{abstract}

Millions of people listen to podcasts, audio stories, and lectures, but editing speech remains tedious and time-consuming.
Creators remove unnecessary words, cut tangential discussions, and even re-record speech to make recordings concise and engaging. 
% \amy{doesn't really clarify the benefits of both approaches (i.e. why would we want them?) --> }
Prior work automatically summarized speech by removing full sentences (extraction), but rigid extraction limits expressivity.
AI tools can summarize then re-synthesize speech (abstraction), but abstraction strips the speaker’s style.
We present \system, a system that flexibly combines extraction and abstraction to condense speech while preserving its content and style.
To edit speech, \system{} first generates possible transcript edits, selects edits to maximize compression, coverage, and audio quality, then uses a speech editing model to translate transcript edits into audio edits.
\system’ interface provides creators control over automated edits by separating low-level wording edits (via the compression pane) from major content edits (via the outline pane).
\system{} achieves higher coverage and removes more speech errors than a state-of-the-art extractive approach.
A comparison study (N=12) showed that \system{} significantly
decreased cognitive load and editing effort in speech editing.
We further demonstrate \system's potential in an exploratory study (N=3) where creators edited their own speech.
\enlargethispage*{12pt}\end{abstract}

\begin{CCSXML}
<ccs2012>
   <concept>
       <concept_id>10003120.10003121.10003129</concept_id>
       <concept_desc>Human-centered computing~Interactive systems and tools</concept_desc>
       <concept_significance>500</concept_significance>
       </concept>
 </ccs2012>
\end{CCSXML}

\ccsdesc[500]{Human-centered computing~Interactive systems and tools}

\keywords{Speech, Audio Editing, Summarization, Creativity Support Tools}

\begin{teaserfigure}
  \includegraphics[width=0.99\textwidth]{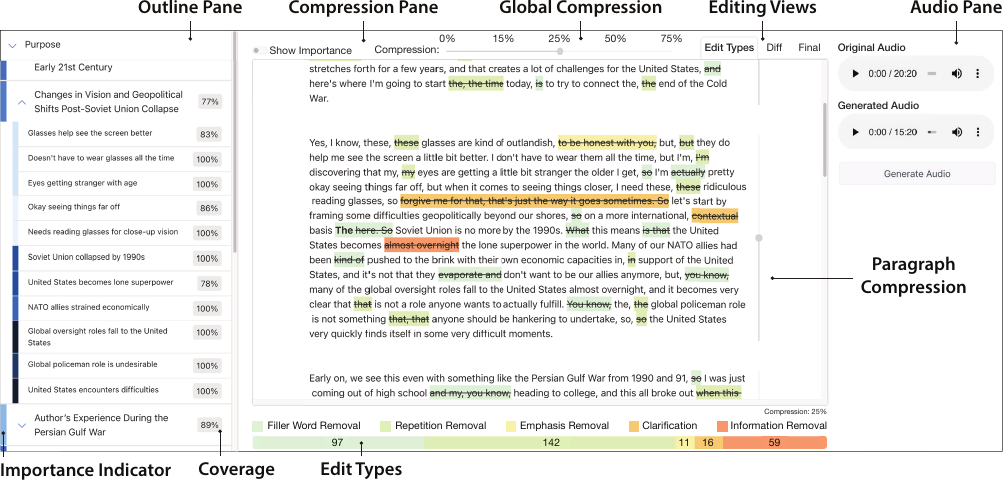}
  \caption{\system{}'s interface lets creators edit their speech by skimming and browsing the outline pane to determine regions to cut, using the compression pane editing the transcript directly or by selecting a global- or paragraph-level compression amount, and listening to the original or generated audio aligned to the transcript in the audio pane.}
  \Description{A screenshot of the TalkLess interface, which is organized into five main panes for editing speech. On the left, the Outline Pane displays a hierarchical structure of topic segments from the transcript, each with a percentage score that indicates how much of the original content is preserved. These segments are grouped under broader themes, such as “Changes in Vision and Geopolitical Shifts Post-Soviet Union Collapse,” and include subtopics like “Glasses help see the screen better” and “NATO allies strained economically.” In the center, the Compression Pane shows the full transcript with color-coded highlights representing different types of edits: filler word removal in gray, repetition removal in green, emphasis removal in yellow, clarification in orange, and information removal in red. Above the transcript is a global compression bar ranging from 0\% to 75\%, allowing users to adjust the overall degree of shortening. To the right of the transcript are editing view tabs labeled Edit Types, Diff, and Final, which let users toggle between different visualizations. On the far right, the Audio Pane provides playback controls for the original and generated audio, as well as a button to generate the edited audio. At the bottom, a horizontal bar chart shows the number of each edit type applied: 97 filler word removals, 142 repetition removals, 14 emphasis removals, 11 clarifications, and 59 information removals.}
  \label{fig:teaser}
\end{teaserfigure}

% wanted to remove this but when I do, the color is messed up
% \received{}
% \received[revised]{}
% \received[accepted]{}

\maketitle

%%%% 01-intro.tex starts here %%%%

\section{Introduction}
% \karim{thesis: people edit their speech as raw speech is long and needs to be shortened}
% people create more and more audio-driven online content (e.g., podcasts, online lectures, interviews, etc..)
% people usually practice their speech so that they can convey their content clearly and it is easier to follow them for the audience.
% However this is not feasible in many contexts. For example when giving very long speeches such as in lectures, or in informal settings such as podcasts and interviews
Millions of people create and consume audio-driven media such as podcasts and lectures.
% While recording speech is easy, editing it 
However, editing speech recordings to be clear, concise, and engaging is tedious and time-consuming~\cite{baume2018contextual}.
For example, podcast creators spend 2-7x the recording length editing the speech (\textit{e.g.}, 7 hours to edit 1 hour of audio~\cite{HighReps2022}).
They carefully cut unnecessary audio including pauses, filler words (\textit{e.g.}, ``um'', ``uh'', ``you know''), repetitions (at a word-, sentence-, or paragraph-level), and irrelevant content, all while making sure to not introduce incoherent or unnatural transitions by adding noticable cuts~\cite{wang2022record}.
Transcript-based editing tools~\cite{rubin2013content,berthouzoz2012tools,Descript,AdobeLabsBlink,alitu, podcastle} let creators edit speech by deleting text or generating new words~\cite{Descript}, but reading raw speech transcripts and deciding what to cut while preserving important information and the speaker’s style (\eg variations in pitch, pacing, and tone to capture listener attention~\cite{rodero2017pitch, smiljanic2009speaking}) remains challenging. 
% cognitively demanding, tedious, and time-consuming task.
Thus, creators without time or resources to edit (\textit{e.g.}, educators, amateur podcasters, or journalists) must publish less polished audio, or choose not to post their work altogether.

Prior automated approaches use \textbf{extractive} speech shortening which extracts segments from the original audio to preserve important content and speaker style~\cite{Descript, pavel2020rescribe, wang2022record}.
For example, ROPE~\cite{wang2022record} selects important transcript sentences, and Rescribe~\cite{pavel2020rescribe} selects words from sentences (\textit{e.g.}, ``An orange cat sat.'' to ``A cat sat.'').
However, rigid extraction in ROPE~\cite{wang2022record} and Rescribe~\cite{pavel2020rescribe}, often preserves repetitions, creates abrupt transitions, and leaves unresolved references (\eg may keep \textit{``she presented the results''} but remove who \textit{``she''} is)~\cite{worledge2024extractive}. 
% (\eg from ``Tina joined the call. She presented the results.'' we may keep ``She presented the results,'' but now it’s unclear who ``she'' is)
% A longstanding solution to achieve fluent summaries with high coverage for text is \textf
% \amy{issues with extraction are really well known/ a longstanding solution is abstractive summarization}
The limitations of extraction are also well explored for text summarization~\cite{worledge2024extractive, ceylan2010quantifying, lin2003potential, zhang2022extractive}.
In response, a longstanding solution for text summarization is to use \textbf{abstractive} summaries that write new sentences and words to summarize sentences and can achieve fluent summaries with high content coverage (\textit{e.g.}, changing \textit{``There is one event in 1990. There is another in 1992.''} to \textit{``There \textbf{were} events in 1990 \textbf{and} 1992.''}). Prior work has explored using abstraction to create text summaries of speech transcripts to ease reading~\cite{li2021hierarchical} or write from speech-to-text~\cite{lin2024rambler}, but prior work has not yet applied abstractive summarization directly to edit speech. One approach may be to create an abstractive speech summary with an LLM, then synthesize the speech (\textit{e.g.}, with ElevenLabs~\cite{ElevenLabs} voice cloning).
But, re-synthesizing speech strips away the speaker's linguistic (\eg word choice) and para-linguistic (\eg emphasis with pitch) style that carry the speaker's identity, message, and delivery~\cite{jm3}.
\revision{
Editors can also combine extractive and abstractive summarization by manually inserting synthesized segments into existing speech using transcript-based editors that allow inserting new words for synthesis through voice cloning~\cite{jin2017, descriptUnderlord, lotus2025, peng2024voicecraft}.
However, manually combining abstractive and extractive summarization is tedious.
}

We present \system{}, a transcript-based editing system to condense speech that contributes a speech summarization algorithm and editing interface to condense speech.
\system's algorithm flexibly combines extraction and abstraction to condense speech by jointly considering the speech transcript and audio through balancing preserving content and speaker style with maintaining coherent and natural-sounding audio.
\system{} transcribes, aligns, and segments the original audio, generates potential transcript edits using an LLM, then selects a set of edits that maximize compression and content coverage without compromising audio quality.
Finally, \system{} translates the selected transcript edits into audio edits and synthesizes transitions between audio cuts and new words.
Editing raw unscripted speech requires many low-level edits to remove filler words and repetitions, and requires high-level edit decisions, as editors often want to intentionally exclude irrelevant content depending on the editing purpose.
\system{} contributes a speech editing interface that separates the two concerns by automating low-level edits to clean up the speech and support editors in making high-level content decisions (Figure~\ref{fig:teaser}).
To make low-level editing more efficient, \system's \textit{Compression Pane} provides dynamic compression controls at both the transcript and segment level to automatically shorten speech, and visualizes filler word and repetition removals to support quick review of low-risk edits.
To support high-level editing, \system{} visualizes information removed by automated edits and indicates content importance based on the editing purpose (\textit{e.g.}, post a class lecture to YouTube) in the \textit{Outline Pane} to help editors identify what content to keep or remove (\textit{e.g.}, a creator's sidenote about his glasses is light blue).

The \textit{Audio Pane} lets creators listen to the original or generated audio. 

We evaluate \system{} with a technical and results evaluation, an authoring user study, and an exploratory user study.
Our technical evaluation reveals that \system{} retains more content and reduces more speech errors than a state-of-the-art extractive audio shortening baseline~\cite{wang2022record}, and participants in our results evaluation significantly preferred listening to \system's shortened speech for 15\% and 25\% compression compared to the baseline.
% Although \system{} edited the audio more heavily than the baseline to retain content, participants in our results evaluation significantly preferred listening to \system's shortened speech for the 15\% and 25\% compression compared to the baseline.
Our user evaluation (N=12) revealed that all editors preferred using \system{} to edit lecture speech recordings to an extractive baseline system, and experienced significantly lower cognitive load as well as significantly higher creative flexibility when editing speech recordings with \system{}.
In our exploratory user study (N=3) creators used \system{} to edit their own speech and compare their results with fully re-synthesized versions. Creators expressed concern about losing vocal authenticity with re-synthesis and preferred \system{} to preserve their original speaking style.

In summary, our work contributes:
\begin{itemize}
    \item An automatic speech shortening algorithm that combines extraction and abstraction to condense speech by jointly considering the speech's transcript and audio.
    \item \system{}, a speech editing interface that supports editors by separating editing concerns into automated low-level edits to clean up speech and information highlighting to leave high-level content edits to the editor.
    % \item A comparison of \system's results with those produced by the state-of-the-art automatic speech shortening method.
    \item A results evaluation and two user studies demonstrating why editors prefer \system's results and its interface for editing others' and their own speech.
\end{itemize}\vspace*{-6pt}

\begin{table*}[t]
\begin{tabular}{@{}p{.5\textwidth} p{.5\textwidth}@{}}
\toprule
\textbf{TalkLess} & \textbf{Prior Extractive Approach~\cite{wang2022record}} \\
\midrule
I was nervous \deletion{at first, like, you know,}because I hadn’t done \deletion{anything like} this before. But I kept going. And \deletion{then, um,} one day it just clicked. \insertion{which} \deletion{That moment} gave me the confidence to keep pushing forward. &
I was nervous at first, like, you know, because I hadn’t done anything like this before.\deletion{But I kept going.}And then, um, one day it just clicked.\deletion{That moment gave me the confidence to keep pushing forward.}\\
\bottomrule
\end{tabular}
\caption{Compression with \system{} (left) vs. a prior extractive approach~\cite{wang2022record} (right). \system{} removes speech errors and condenses content across sentence boundaries, the prior approach leaves speech errors and drops the speaker’s main point.}
\Description{
Comparison of compressed outputs from TalkLess and a prior extractive approach, using the same original transcript. Both versions are shortened to a similar length, but differ in fluency and content preservation. In the TalkLess output, disfluencies like “at first, like, you know,” and “then, um,” are removed, while the sentence remains fluent and structurally intact. A small insertion replaces “That moment” with “which” to improve flow. As a result, TalkLess keeps the speaker’s core message that gaining confidence was a turning point. In contrast, the extractive approach leaves in disfluent phrases like “like, you know” and “then, um,” while cutting the full sentence about gaining confidence. This results in an output that is less coherent and loses the key concluding insight from the speaker.
}
\end{table*}

\section{Background}
\system{} provides a method and interface to shorten speech with a transcript and thus relates to prior work on speech editing tools, text summarization, and automatic speech shortening. \vspace*{-6pt}

\subsection{Transcript-based Editing Tools}
%Timeline-based editing
Traditional audio editing workflows rely on timeline-based editors~\cite{AdobeAudition, Audacity}, where editors manually remove unnecessary filler words, repetitions, and unimportant content by making precise waveform-level cuts.
However, editors must listen closely to the audio to determine what to cut and carefully place cuts so as not to introduce audio errors, which is a time-consuming and tedious task.

% transcript-based editing
To make audio editing easier and more efficient, prior work introduced transcript-based
editing, which allows users to edit audio or video similar to a text document by time-aligning the words in the speech transcript with words in the audio~\cite{Descript, berthouzoz2012tools, casares2002simplifying, fried2019text, huber2019b, rubin2013content, rubin2015capture, shin2016dynamic, leake2024chunkyedit, truong2019tool, rubin2015capture, lotus2025}.
However, reading long, unstructured transcripts remains cognitively demanding as raw speech transcripts include disfluencies and are often unstructured.

% support for finding where to cut
\revision{
Prior systems aim to support editors in surfacing potential edits by highlighting segments with important content~\cite{wang2022record, wang2024podreels, truong2019tool}, narration errors (\eg filler words, unclear pronunciation, bad flow)~\cite{rubin2015capture, Descript}, or grouping thematically coherent transcripts into chunks for further editing~\cite{leake2024chunkyedit}.
For example, Truong~\etal~\cite{truong2019tool} highlight memorable moments in social conversation transcripts, and ROPE~\cite{wang2022record} highlights relevant sentences to include in the final edited audio to support editors in preserving important content.
However, while these systems assist editors in locating content to preserve or speech to remove, they either require the user to manually cut their speech~\cite{truong2019tool, leake2024chunkyedit}, which is tedious and time-consuming, or give little to no editing flexibility, for example, by only letting editors toggle sentences to include~\cite{wang2022record}.
While Descript~\cite{Descript} automatically detects and removes filler words and word repetitions, it is limited to a fixed set of filler words, cannot detect semantic content repetitions and content to cut to make the audio more concise.
}

Beyond highlighting edits, prior work also aimed to support users in browsing and skimming long transcripts by providing section summaries~\cite{pavel2014video}, transcript segments based on topics~\cite{fraser2020temporal, truong2021automatic, pavel2014video}, and outlines~\cite{huh2023avscript} which can be used to find content to cut.
Similarly, to make reading and navigating speech transcripts easier, we segment the transcript into semantically meaningful segments and provide an outline based on information included in the speech.
We expand on prior transcript navigation aids and visualize edit types to make skimming and reviewing edits easier.

\subsection{Text Summarization}
\revision{
Text summarization condenses text content while preserving the core meaning, and is widely categorized as either extractive~\cite{gu2021memsum, Zhang2023ExtractiveSV, Liu2019FinetuneBF, filippova2015sentence} or abstractive~\cite{gehrmann2018bottom, gupta2019abstractive, lin2019abstractive, paulus2017deep, nallapati2016abstractive} summarization.
% Extractive summarization preserves original wording verbatim and is popular in reading-focused applications where preserving original phrasing matters~\cite{gu2024ai, chen2023marvista}.
Extractive summarization condenses content by selecting a subset of existing sentences or words from the original text without modification and is popular in reading-focused applications where preserving original phrasing matters~\cite{gu2024ai, chen2023marvista}.
To facilitate text skimming GP-TSM~\cite{gu2024ai} generates multiple levels of extractive summaries through LLM-based sentence compression, then overlays these summaries on top of each other to visually occlude less important words.
Marvista~\cite{chen2023marvista}, a human-AI reading support tool, uses extraction by automatically choosing a summative subset of text for users based on their time budget and questions they want to answer.
However, extraction misses opportunities to improve flow and conciseness~\cite{worledge2024extractive}, especially when dealing with informal, repetitive, or unstructured text like speech transcripts.
In contrast, abstractive summarization can be more flexible, concise, and coherent through the generation of new text~\cite{worledge2024extractive}.
For example, while extractive summarization might select the sentences \textit{``The meeting started at 9 AM. We discussed the budget. The budget was over by 20\%.''}, abstractive summarization generates \textit{``The 9 AM meeting revealed a 20\% budget overrun.''}
Prior tools use abstractive summarization to transform documents~\cite{august2023paper, dang2022BTG}, audio~\cite{wang2022record, zhong2021qmsum, shang2018unsupervised, li2021hierarchical, lin2024rambler}, or video~\cite{lotus2025, wang2024podreels} into text summaries.
Text simplification also uses an abstractive method to replace complex language with simpler alternatives~\cite{zhong2020discourse, Cripwell2023DocumentLevelPF}.
However, abstraction sacrifices original phrasing and style for improved conciseness and coherence~\cite{worledge2024extractive}.
We flexibly combine extractive and abstractive text summarization for audio summarization to condense speech while preserving its original content and style.
% 
%
% However, abstractive summaries contain new text and translating them back into audio requires re-synthesizing the full speech.
% While re-synthesizing based on an abstractive summary through voice cloning~\cite{ElevenLabs, peng2024voicecraft} produces speech with high speaker similarity, it strips linguistic (\eg word choice) and para-linguistic features (\eg volume, pace, tone, pitch, pauses), and thus speaker identity~\cite{jm3}.
% We aim to flexibly combine extraction and abstraction to summarize speech by creating mostly-extractive summaries to prioritize preserving audio speaker and use abstraction only when necessary to make the speech as condense as possible.\karim{maybe the last two sentences can be moved down into the next section?}
}

\subsection{Audio Summarization}
\revision{
Prior work explored abstractive and extractive text summarization for audio summarization to let audience members consume, skim, and navigate audio recordings efficiently.
Prior work explored two main approaches to summarize audio: generating text summaries from audio transcripts~\cite{zhong2021qmsum, shang2018unsupervised, li2021hierarchical} or directly editing audio to create shortened audio output~\cite{wang2022record, lotus2025, Descript, pavel2020rescribe}.
For example, Li and Chen~\etal~\cite{li2021hierarchical} support people quickly skim long audio by first transcribing the recording, then generating high-level text summaries that users can drill into progressively longer and more detailed summaries.
However, abstractive text summaries of audio can only be converted back to audio via full speech re-synthesis~\cite{jin2017, lotus2025}.
For example, Lotus~\cite{lotus2025} transforms long-form videos into short-form videos by transcribing the audio, generating an abstractive transcript summary with an LLM, then synthesizing the text summary using voice cloning~\cite{ElevenLabs, peng2024voicecraft}.
Although full re-synthesis through voice cloning produces speech with high speaker similarity, it strips linguistic (\eg word choice) and para-linguistic features (\eg volume, pace, tone, pitch, pauses), and thus speaker identity~\cite{laserna2014like, jm3}.}

\revision{
Prior automatic approaches that directly edit the audio use extraction to preserve speaker identity by reusing existing speech segments through filler word removal~\cite{Descript}, word-level edits~\cite{pavel2020rescribe}, or sentence-level extraction~\cite{wang2022record}.
Descript~\cite{Descript} automatically removes a set list of filler words (\eg ``um'', ``uh'') from speech recordings.
Rescribe~\cite{pavel2020rescribe} reduces speech length by automatically removing less important sentences and words within sentences while preserving grammar.
ROPE~\cite{wang2022record} extracts speech based on transcript sentences similar to an abstractive transcript summary to preserve core content.
Further, extractive approaches can cause grammatical and audio errors when combining disjoint sentences, abrupt topic transitions, and miss opportunities to shorten across sentences~\cite{worledge2024extractive, furui2004speech, wang2022record}.
Our method directly edits the audio for audio summarization and combines extraction and abstraction to preserve original speaker style while maximizing content condensation by generating mostly-extractive summaries that use abstraction and thus synthesis only when necessary.
}

%%%% sections/02-background.tex ends here %%%%

%%%% 03-system.tex starts here %%%%

\section{\system}
We present \system, a system that supports automatic shortening of raw speech recordings. We designed \system{} according to design goals that we derived from our review of prior work.

\subsection{\system{} Design Goals}
% Transcript-based editing tools such as Descript~\cite{Descript} and Adobe Blink~\cite{AdobeLabsBlink} support professional editors and creators without audio editing experience to edit speech using text.
Creators and editors spend time and effort to edit raw speech recordings to make them clear and engaging.
We aim to design a transcript-based editing system that supports editors and creators in editing raw speech recordings (\eg class recordings, podcast episodes, or interviews) using text to make them more concise while preserving their original content and style: \\ 
% As scripted speech can be engaging on its own (\textit{e.g.}, a carefully scripted address, or a skit), we aim to improve the process for unscripted or lightly scripted speech that includes a range of content from educational lectures and interviews, to casual podcast conversations. We designed our system with the following goals: \\

\noindent \textbf{G1: Support surfacing and removing unnecessary speech. } 
Not all speech carries equal communicative importance. Creators thus remove unnecessary speech, but distinguishing essential from non-essential speech is tedious and time-consuming~\cite{baume2018contextual}. While scripted speech or written text is carefully crafted to convey a message, unscripted or lightly scripted speech frequently contains disfluent speech~\cite{jm3,rubin2013content, kirkland2023pardon} and tangential content~\cite{wang2022record,rubin2013content}. 
% Compared to scripted speech or prepared text that is crafted ahead of time to convey a specific message, unscripted speech frequently contains non-essential segments due to inefficient or disfluent speech constructions~\cite{jm3,rubin2013content} and tangential content~\cite{wang2022record,rubin2013content}. 
% When speaking 
For example, Jurafsky \etal \cite{jm3} reveal common disfluencies: filler words (\eg ``But, \textit{uh}, that was absurd''), word fragments (\eg ``A guy went to a \textit{d-}, a landfill''), repetitions (\eg ``it was just a \textit{change of, change of} location''), and restarts (\eg ``it's \textit{-}, I find it very strange'').
Prior work also showed that creators want to remove tangential content (\textit{e.g.}, a side story, meta discussions) from speech~\cite{wang2022record, rubin2013content}.
% \amy{not actually sure if anyone said this but it would be handy}
We aim to support editors in surfacing and removing unnecessary speech. \\
    % Not all verbal content\amy{not quite sure what this means -- is this about the words?} in speech carries equal informational or communicative importance.
    % Baume \etal describe that speakers clean up their audio by removing irrelevant content or disfluencies but find it challenging to distinguish essential from non-essential speech elements \cite{baume2018contextual}.
    % Jurafsky \etal \cite{jm3} highlight 4 disfluency types: filler words (\eg ``But, \textit{uh}, that was absurd''), word fragments (\eg ``A guy went to a \textit{d-}, a landfill''), repetitions (\eg ``it was just a \textit{change of, change of} location''), and restarts (\eg ``it's \textit{-}, I find it very strange'').
    % Prior work supports this by showing that some speech segments convey core ideas, while others include details that are only tangential to the primary message \cite{wang2022record, rubin2013content}.
    % Our system should provide an automated compression approach that can distinguish between content that can be removed without semantic loss and elements that contribute to the primary message.

\noindent \textbf{G2: Preserve important information.} After editing speech, the speech should retain the core message. We thus aim to achieve high coverage of important information, similar to prior work~\cite{wang2022record}. We prioritize removing speech that is unlikely to change the meaning of the speech and use abstraction to achieve higher coverage for a similar length~\cite{nenkova2011automatic}. We then let creators manually make any substantial changes to the information provided. \\
% After editing speech, the speech should contain the core message it is attempting to convey. A longstanding goal in speech and text compression is to achieve high coverage of the information in the original speech or text~\cite{}. To preserve important information, we aim to achieve high information coverage in our automated compression by prioritizing removal of speech that does not change the information provided (\textit{e.g.}, repetitions, restarts) and using abstraction to gain additional coverage for a similar length. 
% We aim to then support creators to explicitly make any substantial changes to the information provided in the speech (\textit{e.g.}, a lecturer may choose to remove all information about deadlines). \\

\noindent \textbf{G3: Preserve speaker style.}
Speakers vary in how they use linguistic (language use and construction) and para-linguistic (volume, pace, tone, pitch, pauses) variations in their speech~\cite{chambershandbook,jm3}. Speakers consciously and unconsciously use such variations to guide audience attention (\textit{e.g.}, raise volume for a key point)~\cite{giles2012speech, erickson1978howard} and reinforce their identity~\cite{chambershandbook, schuller2013computational} as such styles inform how speakers are perceived~\cite{fast2008personality, sanford_speech_1942, Sapir1927SpeechAA}. 
For example, speakers may pronounce ``something'' as ``somethin'' due to regional ties, frequently use a specific set of words (\textit{e.g.}, ``wonderful'' correlates to agreeable personalities~\cite{YARKONI2010363}), or use selected vocabulary for core concepts (\textit{e.g.}, ``affordance''). We aim to preserve speakers' linguistic and para-linguistic styles while shortening to retain their identity.
\\
% in our abstractive and extractive compression. \\

% \cite{chambershandbook, schuller2013computational}.
%     Studies have shown that linguistic constructions not only represent but also shape how speakers are perceived \cite{fast2008personality, sanford_speech_1942, Sapir1927SpeechAA}.
\noindent \textbf{G4: Support granular control and efficient review of automated edits.} Compared to manual editing, automating speech edits shifts the burden from producing to reviewing the edits. We aim to reduce the effort required to review edits by providing flexibility in the level of edits to review (\textit{e.g.}, global, paragraph, fact), and supporting skimming the edits with visualizations.\\
% We aim to provide multiple level of edit control (\textit{e.g.}, global, paragraph, fact) to support granular production and review of edits rather than forcing users to review all edits at once. We also aim to provide visualizations to lower the cognitive burden when skimming potential edits. \\
% Different communicative scenarios demand different edits: a professor may remove logistical class details while preserving the conceptual content, while a podcast editor may want to maintain the narrative flow without including tangential discussions. Our interface design should therefore provide multiple levels of compression control and allow users to align automatic edits with their specific communicative goals. \\

\noindent \textbf{G5: Avoid audio errors.} Editing speech occasionally introduces distracting audio errors. For example, removing a word risks introducing artifacts (\textit{e.g.}, cutting in the middle of the word, disrupting background noise)~\cite{pavel2020rescribe, wang2022record, furui2004speech} \revision{and the loss of speaker style (\textbf{G3}), for example due to unnatural pacing or missing breaths}. We aim to avoid such audio errors.
\\

\subsection{\system{} Interface}
\system's editing interface consists of a \textit{Compression Pane} that lets creators edit speech wording, an \textit{Outline Pane} that lets creators edit speech by content, and an \textit{Audio Pane} that lets creators preview their results (Figure~\ref{fig:teaser}).
% The \textit{Compression Pane} consists of three views (\textit{Edit Types}, \textit{Diff}, and \textit{Final}) and lets editors perform transcript-based audio edits by removing or inserting text.
% The \textit{Outline Pane} lets editors navigate, locate, and evaluate the speech transcript content based on extracted information and an importance indicator.
% Lastly, the \textit{Compression Selectors} (global and per-paragraph) give the editor access to shorten the transcript using automatic system edits.

\subsubsection{Compression Pane}
In \system's compression pane, editors can dynamically shorten their speech on the transcript-level or paragraph-level across four target compressions (15\%, 25\%, 50\%, and 75\% compression).
Editors choose a compression using the slider on the top to globally shorten the transcript across paragraphs.
Choosing a target compression on sliders next to each paragraph compresses only the respective paragraph.
The two-level compression approach lets editors control what parts of the transcript to compress more or less, depending on their editing purpose (\textbf{G1}).

\begin{figure}[t]
    \centering
    \includegraphics[width=\linewidth]{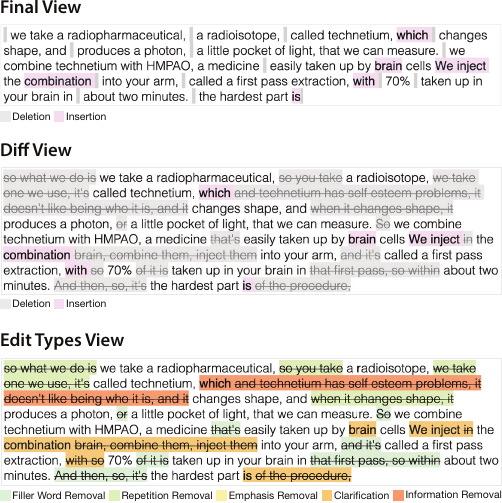}
    \caption{A speech snippet displayed in all three views within the compression pane: The final view displays the final edited transcript with rendered cuts, the diff view renders both \colorbox[HTML]{f6e5f1}{inserted} and {\colorbox[HTML]{eaecec}{\textbf{deleted}}} parts of the transcript, and the edit types view supports skimming and reviewing of automated edits through edit types.}
    \Description{
    Three views of a transcript in the TalkLess interface: Final View, Diff View, and Edit Types View. In the Final View at the top, the edited transcript is displayed with deletions shown in gray strikethrough and insertions shown in purple. This view presents the cleaned and compressed version of the speech. The Diff View in the middle highlights all edits made, showing the full original transcript with deletions and insertions overlaid. Strikethrough text indicates removed phrases, and new text is shown in purple. At the bottom, the Edit Types View shows the same transcript with each edit color-coded by type. Filler word removals are highlighted in gray, repetition removals in green, emphasis removals in yellow, clarifications in orange, and information removals in red. A legend below the view explains the color coding for each edit type. The content of the transcript discusses a radiopharmaceutical process involving technetium and brain imaging, with several edits simplifying technical language and removing asides or informal expressions.
    }
    \label{fig:all-views}
\end{figure}

\system's compression pane features three transcript views to refine and review system edits: \textit{Edit Types}, \textit{Diff}, and \textit{Final} (Figure \ref{fig:all-views}). 
Each view allows manual transcript-based editing on the original speech transcript.
While system edits are accepted by default and rendered in the compression pane, editors can manually refine and override system edits by toggling words they want to keep or remove from the audio, or inserting new words to be synthesised between existing words.
While all views render cuts and insertions made in the audio, each view visualizes system edits at different levels of detail (\textbf{G1, G4, G5}).

% By default the editor shows the \textit{Final} view.
% We designed these editing views to support automatically shortening speech with \system{}.
% Each view is meant to support a different editing goal.

% made in the audio and newly inserted text.
% Across all view, \system{} breaks the transcript apart into paragraphs grouped by content.

While some edits are less risky and can be removed from the speech without significant effect (\ie removal of filler words or repetitions), others are more risky as they may change the meaning or remove content from the original speech (\ie clarifications or information removal)~\cite{jm3}.
The \textbf{edit types view} supports editors in understanding the effect of edits by automatically classifying edits into 5 types:
(ordered by risk level):
\textit{Filler Word Removal} (\textbf{G1}), \textit{Repetition Removal} (\textbf{G1}), \textit{Emphasis Removal} (\textbf{G3}), \textit{Clarification} (\textbf{G3}), and \textit{Information Removal} (\textbf{G4}).
\begin{comment}
\fillword{\textit{Filler Word Removal} (\textbf{G1})}, \repeatword{\textit{Repetition Removal} (\textbf{G1})}, \emphword{\textit{Emphasis Removal} (\textbf{G3})}, \clarify{\textit{Clarification} (\textbf{G3})}, and \info{\textit{Information Removal} (\textbf{G4})}.
\end{comment}
\aptLtoX[graphic=no,type=html]{\colorbox[HTML]{dcefdc}{\textit{Filler Word Removal}}}{\textit{\markerbox{fillerwordcolor}{Filler Word Removal}}} removes unnecessary words or phrases that do not add meaning to the sentence, such as ``um,'' ``like,'' ``you know,'' or ``basically.'' These words are often used as verbal pauses and can be omitted without changing the overall content. \aptLtoX[graphic=no,type=html]{\colorbox[HTML]{e1efc4}{\textit{Repetition Removal}}}{\textit{\markerbox{repetitioncolor}{Repetition Removal}}} removes repeated words, phrases, or ideas that add no new value to the speech to increase clarity and conciseness. \aptLtoX[graphic=no,type=html]{\colorbox[HTML]{faefb7}{\textit{Emphasis Removal}}}{\textit{\markerbox{emphasiscolor}{Emphasis Removal}}} removes words or phrases used to stress or exaggerate a point. These may include words like ``really'' and ``very''. \aptLtoX[graphic=no,type=html]{\colorbox[HTML]{f8ce88}{\textit{Clarification}}}{\textit{\markerbox{clarificationcolor}{Clarification}}} makes an unclear or ambiguous phrase more understandable by providing additional context or rephrasing the phrase. \aptLtoX[graphic=no,type=html]{\colorbox[HTML]{f9a77c}{\textit{Information Removal}}}{\textit{\markerbox{informationcolor}{Information Removal}}} removes details, facts, or sections of the content that are either irrelevant to the main information or not relevant in a different context.

The \textbf{diff view} is similar to the edit types view, but displays cuts in \deletion{gray}and insertions in \insertion{light purple} to highlight the potential audio effect rather than the semantic effect of the edits (\textbf{G5}). 
% We describe their definitions in the next section. \\
% We designed this view to visualize the removed parts (red) of the original speech and newly inserted words (green).
% This supports editors in (re-)evaluating whether the removed part might include relevant information that the editor wants to include in the final speech.
% If editors choose to include previously deleted parts, they can highlight the removed part to include it back into the final transcript. \\

Editors use the \textbf{final view} to read the final transcript to assess the speech flow after the edits occur (\textbf{G4, G5}).
The final view is the default view and displays the final edited transcript, hiding any removed text from the original transcript and indicating cuts in gray (similar to prior transcript-based audio editing tools that display cuts only~\cite{rubin2013content, Descript}). \\
% % after \system{} performs the edits to the audio.
% This view shows only the final edited transcript, including words that will be inserted.
% While the removed content will not be shown in this view, each cut is visualized as a red line within the transcript where the cut will be performed. 

\subsubsection{Outline Pane}
The outline pane lets editors navigate, locate, and edit content included in the original speech transcript (\textbf{G1, G4}).
The outline pane shows content that occurs in the original speech and groups the content into semantically meaningful groups.
Each group is linked to a paragraph in the \textit{Transcript View} and can be further broken down into more fine-granular information that occurs within the paragraph.
Editors can collapse or expand these groups for quick navigation. Hovering over any element in the outline makes the \textit{Transcript View} scroll to and highlight the respective paragraph or aligned low-level outline element. 

To support editors in skimming for content to remove or keep, each element in the information sidebar consists of a blue stripe on the left indicating its importance to the overall speech (\textbf{G1}).
The importance is visualized from least important (light blue) to most important (dark blue).
Editors can guide the importance visualization by typing in a speech purpose at the top of the outline pane. 
For instance, when re-purposing a class lecture to be uploaded on YouTube, the editor may specify ``\textit{lecture uploaded to YouTube for a general audience}'' as the purpose.
The system then re-evaluated each sidebar element's importance according to the specified purpose.
Each sidebar element shows a percentage indicator depicting how many of the keywords of each information element are still included within the final edited transcript to help editors assess what information is being lost as a result of their edits or what information could further be removed (\textbf{G5}).

\subsection{Algorithmic Methods}
\system{} uses a \textit{speech shortening pipeline} that transcribes, aligns, and segments the original speech recording, then generates shortened segment candidates, and after choosing and concatenating the best candidates, automatically edits the audio using an audio editing model.
\system{} \textit{classifies edit types} based on the generated results to make edits easier to skim and review (\textbf{G4}).
Finally, we \textit{extract information} that occurs in the original speech recording and present each bit of essential information together with a computed importance score to the user within the outline sidebar (\textbf{G2}).

\begin{figure*}[t]
    \centering
    \includegraphics[width=\textwidth]{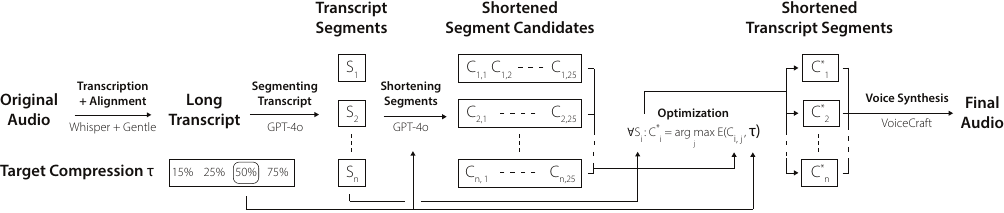}
    \caption{\system{} takes an original audio and a target compression ratio as input. \system{} first transcribes and aligns the audio into a transcript. The long transcript is then segmented into transcript segments $S_1,...,S_n$. For each segment $S_i$, \system{} generates 25 shortened segment candidates $C_{i,1}, ..., C_{i,25}$. For each segment $S_i$, we choose the best candidate $C^*_i$ based on compression, number of edits, edit lengths, and coverage. Finally, \system{} takes the best candidates $C^*_1, ...,C^*_n$ as input to its voice editing model (VoiceCraft~\cite{peng2024voicecraft}) to generate the final shortened audio.}
    \Description{
    The figure shows a pipeline for compressing and synthesizing speech audio from an original recording. It begins with the original audio, which is transcribed and aligned using Whisper and Gentle to produce a long transcript. A target compression rate is selected, such as 15, 25, 50, or 75 percent. The transcript is then split into segments using GPT-4o. Each segment is shortened independently, again using GPT-4o, to generate multiple candidate versions for each segment. An optimization step selects the best shortened version for each segment based on a scoring function that considers both content and the target compression rate. The final selected shortened segments are passed to a voice synthesis system called VoiceCraft, which generates the final audio. The result is a compressed spoken version of the original audio that preserves key content while meeting the target compression rate.
    }
    \label{fig:pipeline}
\end{figure*}

\subsubsection{Automatic Speech Shortening Pipeline}
\label{sec:shortening-pipeline}
Our automatic speech shortening and editing pipeline consists of 2 steps: After transcribing the input audio and aligning each spoken word with the transcript, we create a shortened target transcript during the \textbf{transcript shortening} step, then cut and synthesize based on the transcript to receive the final shortened audio during the \textbf{automatic audio editing} step (Figure~\ref{fig:pipeline}).
\\

\noindent\textbf{Transcript Shortening.}
After pre-processing the audio, we prompt GPT-4o~\cite{openaiapi} to segment the transcript into semantically distinct segments $S_1, S_2, ...S_n$ based on the speech's content. 
Early experiments demonstrated our pipeline works best when compressing smaller segments rather than the full transcript (similar to prior text summarization approaches \cite{gu2024ai, Li2021hierarchicalspeechsummarization}).
% This also reduces lower quality LLM outputs that are more likely to occur with long transcript input context.\amy{<-very vague remove}
% We also experimented with sentence-level segments, however this limits compression and simplifies words and phrases within sentence boundaries. Hence, this does not allow for merging, simplification, and removal of redundancies or repetitions that span across multiple sentences.
% \amy{should we just remove the last two sentences? not clear how they add something?}

For each target compression $\tau \in \{0.15, 0.25, 0.5, 0.75\}$ and segment $S_i$, we generate a set of 25 shortened candidate transcripts $C_{i,1}, C_{i,2}, ...C_{i,25}$ by providing GPT-4o~\cite{openaiapi} with each segment and a prompt that describes eight desired characteristics for shortened transcripts (\ref{sec:appendix_prompt_candidate}) based on our design goals, \eg~removing filler words and repetitions (\textbf{G1}), preserving original information (\textbf{G2}), style (\textbf{G3}), and unique words (\textbf{G3}), limiting word insertions (\textbf{G3}), and not changing word spellings (\textbf{G3, G5}).
% We chose to generate 50 candidates as pilot experiments demonstrated that the number of unique edits introduced for each new candidate does not increase much after 50 candidates.\karim{I have to test whether this is still true for 50 instead of 25 candidates with the new model. Or should we just remove this sentence?}

\revision{
Early experiments revealed that prompting an LLM to generate transcripts that follow our design goals (\textbf{G1-G5}) and target compressions (15\%, 25\%, 50\%, 75\%) produces unreliable results and thus creates an unpredictable user experience (Table \ref{tab:ablation_table}). Hence, \system{} uses an optimization-based approach to generate reliable results.
% When comparing prompting an LLM only with \system's optimization by generating 25 transcripts with each method for 4 audios and all 4 target compressions (15\%, 25\%, 50\%, 75\%), the LLM-only outputs deviated from the compression target by 0-73 percentage points ($\mu=22$, $\sigma=5$). In contrast, \system{} deviated from the compression target by 1-3 percentage points ($\mu=2$, $\mu=1$). LLM-only also risked synthesizing more transcript words, ranging from 0-43\% ($\mu=10\%$, $\sigma=0.4\%$) compared to 0-22\% ($\mu=3\%$, $\sigma=1\%$) with \system{}, and missing more information, ranging from 33\% to 93\% coverage ($\mu=62\%$, $\sigma=1\%$) compared to 65-96\% coverage ($\mu=84\%$, $\sigma=4\%$) with \system{}.
}

To automatically select the optimal candidate for each segment, we score each candidate transcript $C_{i,j}$ based on a target compression $\tau$ using a candidate evaluation function $E(C_{i,j}, \tau)$ that combines four metrics based on our design goals:
\begin{align*}
E(C_{i,j}, \tau) &= \lambda_1 \cdot E_{comp}(C_{i,j}, \tau) + \lambda_2 \cdot E_{edits}(C_{i,j}) \\
     &\quad + \lambda_3 \cdot E_{len}(C_{i,j}) + \lambda_4 \cdot E_{cov}(C_{i,j})
\end{align*}

We empirically determined weights between 0 and 1: $\lambda_1 = 0.4$, $\lambda_2 = 0.15$, $\lambda_3 = 0.1$, $\lambda_4 = 0.35$.
To receive the desired compression level, we compute
\(E_{comp}(C_{i,j}, \tau)\) which is the difference between the candidate and target compression $\tau$, calculated by computing the ratio of the word length of the shortened candidate \( C_{i,j} \) to the word length of the original segment $S_i$:
\begin{align*}
    E_{comp}(C_{i,j}, \tau) = 1 - \lvert \frac{\textit{length}(C_{i,j})}{\text{length}(S_i)} - \tau \rvert
\end{align*}
% We maximize the compression rate within our scoring function to maximize the removal of unnecessary speech (\textbf{G1}).

\revision{For example, for 75\% compression, $E_{cov}(C_{i,j})$ may generate:}

\begin{quote}
The quick \deletion{brown} fox jumped over the \insertion{sleeping}\deletion{lazy} dog \deletion{sleeping quietly} in the \deletion{sunny} garden.
\end{quote}

To minimize audio errors, we compute \(E_{edits}(C_{i,j})\), which is the number of edits required to get from the original audio to the final edited audio (\textbf{G5}).
We use the Needleman-Wunsch~\cite{needleman-wunsch} algorithm to receive a list of edit instructions that describe the differences between the texts of $C_{i,j}$ and $S_i$.
These instructions are then grouped into chunks of consecutive edits, with each chunk representing a single edit operation as performed by the voice editing model.
The total number of edit operations is normalized by the maximum possible number of edits, defined as half the word count (\ie every second word is edited) of the original segment (\eg ``The \deletion{quick} brown \deletion{fox} jumps \deletion{over} the \deletion{lazy} dog'' has at most 4 edits):
\[
E_{edits}(C_{i,j}) = 1 - \frac{\#\textit{ of edits}}{(\textit{length}(S_i)/2)}
\]

\revision{Adding $E_{edits}(C_{i,j})$ now reduces the number of edits required:}

\begin{quote}
The \deletion{quick brown}fox jumped over the \insertion{sleeping}\deletion{lazy} dog \deletion{sleeping quietly} in the sunny garden.
\end{quote}

We avoid long speech synthesis that risks stripping speaker identity (\textbf{G3}) and introducing artifacts (\textbf{G5}), by computing \(E_{len}({C_{i,j})}\) which is the mean length of insertions within a segment \( C_{i,j} \):
\[
E_{len}(C_{i,j}) = 1 - \frac{\sum (\textit{length of insertions})}{\#\textit{ of insertions}}
\]

\revision{Adding $E_{len}({C_{i,j})}$ now removes the insertion:}

\begin{quote}
The \deletion{quick brown}fox jumped over the lazy dog \deletion{sleeping quietly} in the sunny garden.
\end{quote}

To maximize preserving important information (\textbf{G2}), we compute \(E_{cov}(C_{i,j})\) by matching each sentence $s \in  S_i$ to the most similar sentence $c \in C_{i,j}$, using a sentence transformer model (all-mpnet-base-v2~\cite{all-mpnet-base-v2}), and then averaging the best matches to get a final coverage score:
\[
E_{cov}(C_{i,j}) = \frac{1}{|S_i|}\sum_{s \in S_i}\max_{c \in C_{i,j}}\textit{sim}(s,c)
\]

\revision{After adding $E_{cov}(C_{i,j})$, the information that the dog was asleep is now included:}

\begin{quote}
The \deletion{quick brown}fox jumped over the \deletion{lazy} dog sleeping \deletion{quietly} in the sunny garden.
\end{quote}

Finally, for each segment \( S_i \), we select the candidate with the highest score $C_i^{*} = \arg\max_j E(C_{i,j})$ and concatenate all \( C_i^{*} \) to construct the final shortened transcript.
\\

\noindent\textbf{Automatic Audio Editing.}
We generate the final shortened audio based on the original audio and the shortened transcript.
We use the list of edit instructions received from applying the Needleman-Wunsch algorithm~\cite{needleman-wunsch} between the original transcript and the final shortened transcript.
\system{} supports three types of edits: (1) \textbf{deletions}, or removing segments from the original audio, (2) \textbf{insertions}, or adding new segments to the original audio, and (3) \textbf{replacements}, or replacing existing segments with new ones.

For every \textbf{deletion} (\textit{``...to figure out {\deletion{how to weather or}} how to maintain...}),
 we identify the start and end timestamps of the segment to remove and cut it from the audio.
To avoid noticeable cuts in the audio, we synthesize a smooth cut using VoiceCraft~\cite{peng2024voicecraft}, an open-source speech editing model.
We use the 10 seconds of the original audio that surround the edit as input to the model, as we found this is the ideal length to match the speaker's original speech characteristics and not fall out of the model's training distribution.
Because natural speech blends adjacent words, we re-synthesize half of the word before and after the cut.
To reduce synthesis artifacts, we align transition start points with natural pauses when possible, and choose transitions shorter than 0.6 seconds, or the shortest among 5 generations, whichever is first (\textbf{G5}).

We synthesize each \textbf{insertion} (\textit{``...everyone \insertion{has} to sort of make things work....''}), 
using the same approach as generating transitions for deletions.
To avoid bad generations, we predict the expected generation length based on the mean phoneme length surrounding the edit, then choose the closest after generating 5 generations.

We define \textbf{replacements} as insertions directly followed by deletions (\textit{``...books, but \insertion{try }\deletion{they're trying} to send money...''}) and treat them like insertions after removing the deleted part from the audio.
Each edit generation and the preceding original transcript segment that remains unchanged are appended to the previous edit output, gradually creating the final shortened audio.

\subsubsection{Classification of Edit Types}
For each segment and compression level, we use the list of required edits received by applying the Needleman-Wunsch algorithm~\cite{needleman-wunsch} within the Speech Shortening Pipeline (Section~\ref{sec:shortening-pipeline}), to classify each of the five edit types (\textbf{G4}).
For each edit, we use GPT-4o~\cite{openaiapi} to classify its edit type by prompting it with the text before and after the edit and definitions of edit types (see prompt in ~\ref{sec:appendix_prompt_edit-types}).

\subsubsection{Extracting Information and Evaluating their Importance}
For each segment of the transcript, we extract the atomic pieces of information into bullet points using a GPT-4o prompt (see ~\ref{sec:appendix_prompt_content-extraction}).
This approach is inspired by prior work on automatic outline generation \cite{ghazimatin2024podtile} and atomic fact extraction \cite{tang2024minicheck, kriman2024measuring}.
We create information groups by generating a summary of all information pieces included within a transcript segment.
For each information piece, we evaluate its importance by prompting GPT-4o to generate an importance score between 1 and 10 based on the information piece, the original transcript, and a user-specified purpose (\textbf{G2}).
This prompt can be found in ~\ref{sec:appendix_prompt_importances}.
As LLM-generated scoring tends to mostly output scores in the range of 7 to 10, we redistribute the scores to be between 1 and 10.
Each information group is assigned the mean importance score of all the information pieces it includes.

\subsection{Transcript Shortening Ablation}
\revision{
% We compared and analyzed shortened transcripts (\ie target compression deviation, portion of words synthesized, and coverage) generated by an LLM only without optimization using our prompt (see \ref{sec:appendix_prompt_candidate}) and \system's optimization-based approach.
% We compared and analyzed shortened transcripts (\ie target compression deviation, portion of words synthesized, and coverage) generated by two methods using the same prompt (see~\ref{sec:appendix_prompt_candidate}): (1) using an LLM only and (2) adding \system's optimization-based approach.\\
We compared and analyzed shortened transcripts (\ie target compression deviation, portion of words synthesized, and coverage) generated by two methods: (1) using an LLM only with our prompt (see~\ref{sec:appendix_prompt_candidate}) and (2) using our same LLM prompt combined with our optimization-based approach.
% We compared shortened transcripts (\ie target compression deviation, portion of words synthesized, and coverage) both generated from our LLM prompt (see~\ref{sec:appendix_prompt_candidate}), with and without \system's optimization.\\
We generated 25 transcripts with each method for 4 audio files and all 4 target compressions (15\%, 25\%, 50\%, 75\%).
Our experiment revealed that the LLM-only outputs deviated from the compression target by 0-73 percentage points ($\mu=22$, $\sigma=5$). In contrast, \system{} deviated from the compression target by 1-3 percentage points ($\mu=2$, $\mu=1$). LLM-only also risked synthesizing more transcript words, ranging from 0-43\% ($\mu=10\%$, $\sigma=0.4\%$) compared to 0-22\% ($\mu=3\%$, $\sigma=1\%$) with \system{}, and missing more information, ranging from 33\% to 93\% coverage ($\mu=62\%$, $\sigma=1\%$) compared to 65-96\% coverage ($\mu=84\%$, $\sigma=4\%$) with \system{} (Table~\ref{tab:ablation_table}).
}

\subsection{Implementation}
We implemented a frontend and a backend for \system{}.
The frontend was implemented using \textit{React.js}.
The backend consists of a Python Flask API, which contains most of the pipeline steps and handles their procedural flow.
We used rev.ai~\cite{revai} and Gentle forced-alignment~\cite{gentle} to transcribe and align the input audio.
We also hosted the open-sourced VoiceCraft \cite{peng2024voicecraft} speech editing model on a server with an A40 GPU (48 GB VRAM).

%%%% sections/03-system.tex ends here %%%%

%%%% 04-evaluation.tex starts here %%%%

\section{Technical \& Results Evaluation}
We evaluated the shortened audio results generated by \system{}.
We compared our results with those shortened using ROPE \cite{wang2022record}, serving as a baseline of previous automatic audio shortening methods.

\begin{table}
\centering
\resizebox{3.33in}{!}{%
\begin{tabular}{lp{7.5cm}} 
\toprule
\textbf{Evaluation Metric} & \textbf{Description} \\ 
\midrule
\textsc{Speech Disfluencies} & The combined count of \textit{filler words} (e.g., “uh”, “um”) and \textit{repetitions} (e.g., “[..] in my, in my perspective [..]”), counted using a speech transcript linked to the audio~\cite{Descript}.\\ 
\hline
\textsc{Coherence Errors} & Edit-induced errors that disrupt natural flow (e.g., abrupt transitions through cuts in the middle of a word) and inconsistencies in speech (e.g., missing co-references), counted by listening to the audio. \\ 
\hline
\textsc{Coverage} & Percentage of content from the original transcript retained in the shortened version, measured by computing cosine similarities between sentence embeddings of the transcript and summary, selecting the highest similarity per transcript sentence, then averaging these scores.\\ 
\hline
\textsc{Style Preservation} & Cosine similarity between content-independent style representations~\cite{wegmann2022same} for transcripts generated by \system{}, the baseline, and synthesized abstractive summaries.\\ 
\hline
\textsc{Audio Preference} & Human annotator’s preference for one audio sample over another. \\ 
\bottomrule
\end{tabular}}
\caption{We use six evaluation metrics to evaluate results generated by \system{} and the baseline across all target compressions (15\%, 25\%, 50\%, 75\%).}
\label{tab:metrics}
\end{table}

\subsection{Method}
We selected a total of 4 speech recordings, across 3 categories: podcast, interview, and lecture.
% We recruited a professional with prior experience in audio editing to shorten the three audio files using Descript for golden baselines.
We selected the state-of-the-art approach for extractive audio summarization, ROPE~\cite{wang2022record}, as our baseline. We re-implemented ROPE as it is not open-source and replaced ROPE's abstractive summarization pipeline step with state-of-the-art abstractive summarization through GPT-4o.
% to make shortened speeches comparable.

% We expand on error categorizations used by Subbiah~\etal~\cite{subbiah2024reading} to evaluate short story summaries\amy{I'm curious how ours are similar or different?}, to define
We used five evaluation metrics (Table \ref{tab:metrics}) to evaluate results generated by~\system{} and the baseline for all target compressions (15\%, 25\%, 50\%, 75\%): \textsc{Speech Disfluencies}, \textsc{Coherence Errors}, \textsc{Coverage}, \textsc{Style Preservation}, and \textsc{Audio Preference}. We compute compression rates based on the number of words ratio between the original and the shortened transcripts (\ie a 20\% compression means that the text has been shortened by 20\%). We also include the original audio for \textsc{Speech Disfluencies} and \textsc{Coherence} measures.
To measure how well \system{} and the baseline preserve linguistic style compared to full abstraction combined with voice cloning, we include a synthesized abstractive summary for each compression to evaluate \textsc{Style Preservation}.
We generate abstractive summaries by prompting GPT-4o~\cite{openaiapi} to shorten the transcript to the same target compression ratios without the desired characteristics used in \system{}'s approach.

We recruited 12 human evaluators from our organization to rate their \textsc{Audio Preference} between results generated by \system{} and the baseline based on all 4 original audio files in an online survey.
We counterbalanced the order of comparisons among all evaluators who were unaware of the condition.
% We used the Needleman-Wunsch algorithm and clustered single edits into chunks of consecutive edits. The \textsc{Edit Count} is the number of chunks remaining after clustering.
% We used embeddings produced by the model introduced by Wegmann~\etal to measure how well the author's style is preserved.
We analyzed all results using pairwise t-tests and repeated-measures ANOVAs.

% To evaluate the accuracy of our detection algorithm whether an information bit is included or excluded in the current short transcript, one researcher manually counted the number of correctly and incorrectly classified information bits across all three audio files.

% \subsection{Procedure}
% Human evaluators rank their overall preference between results generated by \system{} and ROPE in a online survey.
% Evaluators were prompted to consider quality of the audio, content, and ease of listening when rating their overall preferences
% Each evaluator performed this task for 6 different audio samples.

% \begin{figure}[t]
%     \centering
%     \includegraphics[width=\linewidth]{figures/ResultsEvaluationGraphs.pdf}
%     \caption{coverage, speech disfluencies and coherence errors}
%     \label{fig:results_eval_graphs}
% \end{figure}
\begin{figure*}[t]
    \centering
    \includegraphics[width=6.69in]{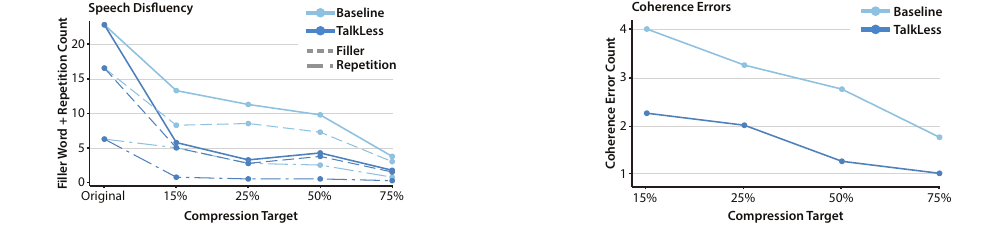}
    \caption{\textsc{Speech Disfluencies} and \textsc{Coherence Errors}. Dotted lines represent filler words and repetition counts.}
    \Description{The figure contains two line charts comparing speech disfluency and incoherence across different compression targets for two systems: the baseline and TalkLess. The left chart shows speech disfluency, measured as the combined count of filler words and repetitions. The x-axis represents increasing compression targets from the original audio to 15\%, 25\%, 50\%, and 75\%. Both systems show a significant reduction in disfluencies as compression increases. TalkLess consistently results in fewer disfluencies than the baseline across all compression levels. The chart also breaks down the disfluencies into filler words and repetitions, with TalkLess exhibiting particularly low counts for both components. The right chart presents incoherence count across compression levels. The baseline shows a steady decline in incoherence as compression increases, starting from a count of 4 at 15\% compression and dropping to below 2 at 75\%. TalkLess consistently maintains lower incoherence than the baseline at every compression level, indicating smoother and more coherent speech output.}
    \label{fig:results_eval_graphs}
\end{figure*}

\begin{figure}[t]
    \centering
    \includegraphics[width=2.5in]{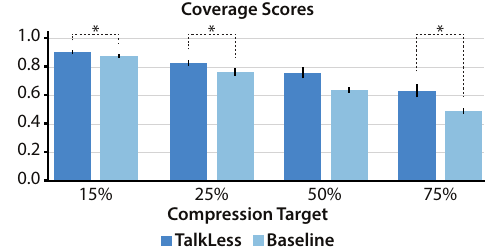}
    \caption{Comparison between coverage for each compression level for \system{} and the baseline. Error bars represent a 95\% confidence interval. *\alphaval{<}{.05}}
    \Description{The bar chart shows coverage scores for TalkLess and a baseline system across four compression targets: 15\%, 25\%, 50\%, and 75\%. Across all compression levels, TalkLess achieves higher coverage scores than the baseline. The differences are most pronounced and statistically significant at 15\%, 25\%, and 75\%, as indicated by asterisks above the paired bars.}
    \label{fig:results_eval_coverage}
\end{figure}

\begin{figure}[t]
    \centering
    \includegraphics[width=2.5in]{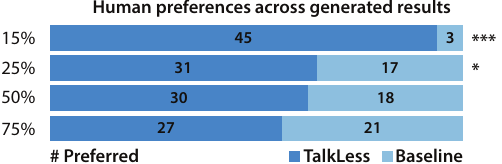}
    \caption{Human preferences. The numbers represent the number of times a method is preferred for each compression. *\alphaval{<}{.05}, ***\alphaval{<}{.001}} 
    \Description{The figure is a horizontal bar chart showing human preferences between TalkLess and a baseline system at four compression targets: 15\%, 25\%, 50\%, and 75\%. Each row represents a different compression target and is divided into two colored segments: dark blue for TalkLess and light blue for the baseline. The number of preferences is labeled inside each segment. At the 15\% compression target, TalkLess received 45 preferences and the baseline received 3. At 25\%, TalkLess received 31 preferences and the baseline received 17. At 50\%, TalkLess received 30 preferences and the baseline received 18. At 75\%, TalkLess received 27 preferences and the baseline received 21. Asterisks appear to the right of the bars for 15\%, 25\%, and 75\%, indicating levels of statistical significance. Three asterisks are shown for 15\%, and one asterisk for 25\%}
    \label{fig:preference_by_compression}
\end{figure}\vspace*{-6pt}

\subsection{Results}
Overall, shortened results generated by \system{} had significantly higher coverage (15\%, 25\%, 75\%), fewer speech disfluencies and coherence errors, and were significantly more preferred (15\%, 25\%) than the baseline.
% Human evaluators significantly preferred shortened audio generated by \system{} over the baseline for 15\% and 25\% target compressions.
Across all conditions, \system{} generated audio with a mean deviation of 3\% ($\sigma = 3.66\%$) from the specified target compression rate. \revision{
Generated results contained synthesized insertions with a mean length of 1.45 words ($\sigma=0.03$) and no insertion was longer than 5 words.
}
% Audio produced by \system{} contains fewer speech disfluencies, regardless of the compression rate (in particular, repetition is suppressed to almost zero for all compression rates) than the baseline (Figure~\ref{fig:results_eval_graphs}). The baseline has more than twice as many disfluencies as \system{} when the compression rate is low, and only reaches a comparable count when the compression rate is very high.
% The chart on the right of figure~\ref{fig:results_eval_graphs} shows that the baseline's results include more coherence errors than \system's results across all compression rates.
\system's results contained fewer speech disfluencies than those of the baseline across all target compressions (Figure~\ref{fig:results_eval_graphs}).
The baseline results contain more than twice as many disfluencies as \system's when the compression rate is low, and only reach a comparable count when the compression rate is very high.
This is because \system's word-level compression allows targeted removal of speech disfluencies while sentence-level compression in the baseline preserves speech disfluencies in extracted sentences.
The baseline produced significantly more coherence errors than \system{} across all target compressions (\alphaval{<}{.01}, Figure~\ref{fig:results_eval_graphs} right).
% We see a higher incoherence count when the compression rate is low, indicating the negative impact of frequently cutting and concatenating audio segments, while our approach keeps the incoherence errors at a low level regardless of compression rate because of synthesizing transitions at the cut points to smooth edits.

Figure~\ref{fig:results_eval_coverage} shows that results generated by \system{} covered significantly more information of the source recording than the baseline for 15\%, 25\%, and 75\% target compression rates (\alphaval{<}{.05}).

% \revision{We found that across all compression levels, there was a significantly larger difference in the number of edits in results generated by \system{} compared to those generated by the baseline ($\mu = 35.04$, $\sigma = 13.22$ vs. $\mu = 6.96$, $\sigma = 2.2$; $t(23) = 11.1$, \alphaval{<}{.001}).} \amy{so what?}

\system's results were significantly more preferred than those of the baseline for 15\% and 25\% target compressions (\alphaval{<}{.05}, Figure~\ref{fig:preference_by_compression})
\system's results were more preferred than those of the baseline for higher compressions, yet there was no significant difference.
Hence, our approach is more suitable for information-rich scenarios where one needs to have a low target compression between 0\% and 25\% to retain most of the information in the original speech.
Investigating preferences for individual audios revealed that \system{} is preferred more often when the original audio contains more disfluencies, incoherence errors, and is more informal.
%\amy{can we share any data on this? e.g., breakdown per video or something?}.
Thus, \system{} may be more useful for spontaneous and unscripted speech.
% \amy{I wonder if this is similar to what we were saying earlier with information-rich scenarios, not quite sure}

We found a significant difference in \textsc{METHOD} ($F_{1.07, 5.37}$, \alphaval{<}{.01}) on text style cosine similarities between transcripts generated by each method and the original transcript across compression targets (Figure~\ref{fig:text-author-similarity}).
Across all compression targets, \system's (\alphaval{<}{.05}) and the baseline's (\alphaval{<}{.01}) transcript style were significantly closer to the original transcript style than abstractive transcripts. \vspace*{-6pt}

\section{User Evaluation}
We evaluated \system's editing interface to investigate its ability to facilitate shortening speech recordings.\vspace*{-6pt}
% We conducted a user study with 8 participants to understand: (1) Can users shorten speech faster using \system{} than when using traditional methods?'' and (2) ``Can users create better shortened speech than when using traditional method?''

\subsection{Method}

\subsubsection{Participants} We recruited 12 participants (8 male and 4 female, ages 21 to 39) from within our organization and from outside using email lists.
All participants had prior experience in audio and/or video editing ($\mu = 3.42$ years, $\sigma = 2.64$ years), and 4 (P2, P5, P6, and P11) had professional experience in audio editing.

\subsubsection{Baseline and study design}
We conducted a within-subjects study where we compared shortening raw speech recordings with \system's interface to a baseline interface.
% We used the state-of-the-art extractive audio summarization approach (\ie ROPE \cite{wang2022record}) for the baseline interface.
The baseline interface is a restricted version of \system's interface where editors can only choose from global compressions generated by the state-of-the-art extractive audio summarization approach (\ie ROPE \cite{wang2022record}) and do manual refinements in the \textit{diff view} and \textit{final view} only (Figure~\ref{fig:baseline_interface}).
\revision{
% We used the most recent extractive audio summarization approach (ROPE~\cite{wang2022record}) and updated the algorithm with GPT-4-2024.
ROPE provides a strong baseline as it extracts important moments, encourages coherence with clause-level edits, and follows compression targets.
We also evaluated an extractive approach using an LLM that used our base prompt (\ref{sec:appendix_prompt_candidate}) but modified to enforce strict extraction with no inserted or modified words.
However, unlike ROPE, LLM-only extraction was unreliable as it did not adhere to extraction constraints or target compressions (similar to LLM-only in Table~\ref{tab:ablation_table}).
We updated ROPE's abstractive summary generation step to use GPT-4~\cite{openaiapi} and modified ROPE's interface to allow word-level edits and dynamic compression between the same target compression rates as in \system{} (Figure \ref{fig:baseline_interface}).
}
% This decision is based on the sentence editor functionality introduced in ROPE\amy{I don't understand this sentence}.
% However, we expand on the functionality by providing fine-grained editing control through word-level editing similar to \system's interface\amy{this is confusing but i think we mean that we allow word-level edits}.
% While ROPE requires users to specify a target duration, we provide users with the same set of target compression rates as in \system{}.
The study has 4 conditions (\textsc{\system{}} vs. \textsc{Baseline}) x (\textsc{Audio 1} vs. \textsc{Audio 2}).
For \textsc{Audio 1} and \textsc{Audio 2}, we selected two history lecture recordings from the same speaker with similar length, as those can be consumed without visuals and contain raw unedited speech~\cite{americanhistory1, americanhistory2}.
We counterbalanced conditions among all participants to alleviate ordering effects.

\subsubsection{Procedure}
During the study, participants completed a demographic profile questionnaire and a consent form (5 minutes).
We then introduced the study and gave participants a tutorial for both interfaces (20 minutes).
The participants complete two editing tasks (2 x 20 minutes = 40 minutes).
After each editing task, participants completed a questionnaire.
The between-task questionnaire includes Likert rating scales extracted from the NASA TLX \cite{NASA_TLX}, the Creativity Support Index (CSI) \cite{cherry14}, and the System Usability Scale (SUS) \cite{brooke1996sus}.
After completion, we interviewed participants (30 minutes).
Participants were compensated with \$30 via PayPal or Venmo and each study session lasted approximately 90 minutes.

\subsubsection{Analysis}
We analyzed task load, creativity support, and usability ratings using paired Wilcoxon Signed Rank tests.
We apply Greenhouse-Geisser correction when the equal-variances assumption is violated (Mauchly’s test \alphaval{<}{.05}).
% When needed, we performed pairwise post-hoc tests with Bonferroni adjustment.
We analyzed participants' interaction data using multiple repeated-measures ANOVAs.

\subsection{Results}
% \amy{Main issue with the user study section is that it feels like it repeats itself several times! It is hard to follow because there are several sections talking about the interactions and the sections aren't unified or clearly differentiated. I wonder if we need to have an overall interactions part then break down other stuff by feature or something}

% Participants highlighted how \system{} made their process ``a lot easier and streamlined'' (P7).
% Compared to the baseline interface, editing with \system{} requires significantly less cognitive load.
All participants preferred \system{} over the baseline interface and expressed enthusiasm for using it in the future to polish audio to make podcasts and recorded presentations easier to follow (P1, P4, P7, P10), re-purpose audio to adapt it to new audiences (P3, P4, P8), or summarize audio to create concise summaries of lengthy lecture recordings or information-rich content (P4, P7).
Participants noted that \system{} saved time and effort, allowing for an efficient workflow of \textit{``first record, then edit to ideal transcript''} (P3). Non-professionals also saw value in \system{} for improving clarity when editing their own recorded speech (P3, P7, P10).

%Participants felt less supported in the baseline interface and new editing strategies emerged when using \system's interface.

% \subsubsection{Future Tool Use}
% \revision{
% P3 notes how \system{} saves them time and effort as it allows them to \textit{``first record, then edit to ideal transcript''}.
% Participants noted polishing audio (P1, P4, P7, \eg editing podcasts or recorded presentations to make them easier to follow), re-purposing audio (P3, P4, P8, \eg to a different audience), and summarizing audio (P4, P7, \eg when creating a video summary with underlying audio transcript) as most exciting use cases where they would like to use \system{} in the future.
% \system{} was seen as particularly helpful editing podcasts (P2, P5, P6, P11) but also for non-professionals who for instance record their own presentations (P3, P7, P10).
% }

\subsubsection{User Interactions \& Subjective Ratings}
\begin{figure}[t]
    \centering
    \includegraphics[width=3.33in]{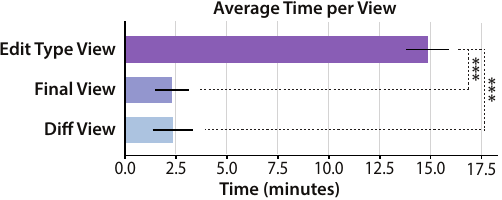}
    \caption{Users spent significantly more time on the Edit Types view than on the Final and Diff view when working with \system{}. ***\alphaval{<}{.001}}
    \Description{The bar chart displays the average time spent per view type in minutes across three conditions: Edit Type View, Final View, and Diff View. The x-axis represents time in minutes, ranging from 0 to approximately 17.5 minutes. Edit Type View, represented by a long purple bar, shows the highest average time of around 15 minutes. Final View and Diff View, shown as shorter blue bars, both have average times of about 2.5 minutes. Error bars are included for all three conditions. Asterisks next to the dotted lines comparing Edit Type View with the other two views indicate statistically significant differences. There are three asterisks in each comparison, denoting a high level of statistical significance.}
    \label{fig:time-spent}
\end{figure}

Participants had similar compression rates with both methods ($\mu$ = 0.58, $\sigma$ = 0.11 with \system{} compared to $\mu$ = 0.58, $\sigma$ = 0.15 with the baseline).
However, participants manually edited significantly fewer words in \system{} ($\mu$ = 477.55, $\sigma$ = 394.58) than in the baseline interface ($\mu$ = 807.45, $\sigma$ = 352.75)(\alphaval{<}{.05}).
Participants frequently switched between views in \system{} to verify and perform edits (Figure~\ref{fig:event_visualizations}).
While the diff view was initially active in \system{}, all but one participant switched to the edit type view within the first third of the session.
Near the end of the study, 5 participants switched to the final view to verify their final edited transcript.
Although participants started on the diff view, they spent significantly more time on the edit types view ($\mu = 12.88$, $\sigma = 2.92$) than on both the diff view ($\mu = 2.35$, $\sigma = 3.28$; \alphaval{<}{.001}) and final view ($\mu = 1.35$, $\sigma = 2.17$; \alphaval{<}{.001}) (Figure~\ref{fig:time-spent}).

\begin{figure*}[t]
    \centering
    \includegraphics[width=\linewidth]{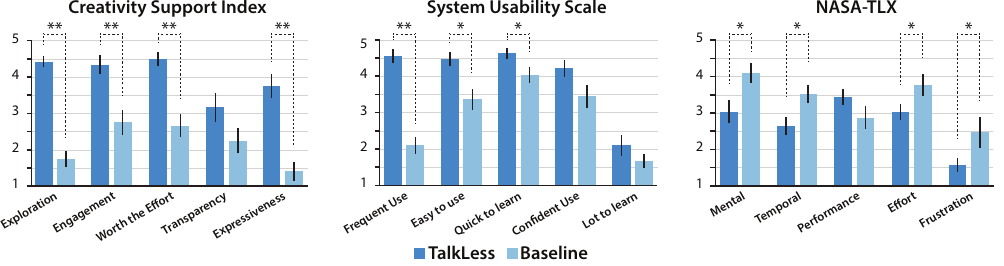}
    \caption{Participant ratings for questions selected from Creativity Support Index, System Usability Scale, and NASA TLX for \system{} and baseline. Error bars represent a 95\% confidence interval. *\alphaval{<}{.05}, **\alphaval{<}{.01}}
    \Description{The figure shows three grouped bar charts comparing TalkLess and a baseline system across multiple subjective evaluation metrics. The first chart, titled "Creativity Support Index," displays five dimensions: Exploration, Engagement, Worth the Effort, Transparency, and Expressiveness. In each case, TalkLess (dark blue) has higher scores than the baseline (light blue). Statistically significant differences are indicated with asterisks above the bars for Exploration, Engagement, Worth the Effort, and Expressiveness, each marked with two asterisks. Transparency is also marked with two asterisks. The second chart, titled "System Usability Scale," includes six categories: Frequent Use, Easy to use, Quick to learn, Confident Use, and Lot to learn. TalkLess has higher scores in all categories except for "Lot to learn," where lower values are better. Asterisks indicate statistical significance for Frequent Use (two asterisks), Easy to use, Quick to learn, and Confident Use (each with one asterisk). The third chart, titled "NASA-TLX," presents five workload-related dimensions: Mental, Temporal, Performance, Effort, and Frustration. For Mental, Temporal, Effort, and Frustration, TalkLess shows lower scores than the baseline. Performance scores are similar between the two systems. Asterisks indicate statistically significant differences in the Mental, Temporal, Effort, and Frustration categories, each with one asterisk.}
    \label{fig:nasa-tlx}
\end{figure*}

Participants reported significantly lower mental demand, temporal demand, effort, and frustration when working with \system{} compared to the baseline interface (\alphaval{<}{.05}, Figure~\ref{fig:nasa-tlx} left).
% We further observed \revision{non-significant} higher performance ($\mu = 3.63, \sigma = 0.74$ vs. $\mu = 2.75, \sigma = 1.16$), lower temporal ($\mu$ = 2.50, $\sigma$ = 0.53 vs. $\mu = 3.86, \sigma = 0.83$) demand and less frustration ($\mu = 1.50, \sigma = 0.53$ vs. $\mu = 2.75, \sigma = 1.28$) when using \system{}.
Using \system, participants found it significantly easier to explore many ideas, felt more engaged in the activity, felt that what they were able to create was more worth their effort, and were significantly more expressive than in the baseline interface (\alphaval{<}{.05}, Figure~\ref{fig:nasa-tlx} middle).
% \system{} helped participants to concentrate non-significantly more on the activity than the baseline interface ($\mu = 3.50, \sigma = 1.07$ vs. $\mu = 2.75, \sigma = 1.16$).
Participants rated \system{} to be significantly easier to use and faster to learn than the baseline (\alphaval{<}{.05}, Figure~\ref{fig:nasa-tlx} right). Participants want to use \system{} significantly more frequently in the future (\alphaval{<}{.01}). We report the means and standard deviations for task load, creativity support, and usability ratings in Table~\ref{tab:ratings_significances}.
% \amy{we're missing a bunch of numbers here (e.g., we usually means and STD and whatnot) -- i'm ok with this but i'd be more okay with it if we had a table in an appendix with all the info}

% \revision{Participants} were non-significantly more confident when using it ($\mu = 4.25, \sigma = 1.04$ vs. $\mu = 3.63, \sigma = 1.19$) compared to the baseline interface. However, participants noted that \system{} requires slightly more to learn to get going ($\mu = 1.75, \sigma = 0.71$ vs. $\mu = 1.50, \sigma = 0.53$).

\subsubsection{\system{} vs. Baseline Compression}
All participants preferred the compression of \system{} over the compression of the baseline.
For instance, due to \system' ability to \textit{``connect sentences''} (P4) or \textit{``make it more concise''} (P12).
P3 preferred \system's compression because it is \textit{``very common that half [of a sentence] is useful and the other half is not.''} 
% For instance, participants highlighted that \system's compression approach is \textit{``great to connect sentences.''} (P4) and \textit{``helps making it [the content] more concise''} (P12).
Participants found the initial compression levels in the baseline interface less helpful and all participants wished for more flexible edits in the baseline interface, as 
% P4 commented on the baseline compressions, \textit{``The system excluded many sentences here, but it's still very relevant... I am thinking how I can revert it back,''} and found it \textit{``hard to notice all filler words and delete them.''}
P4 described:
\begin{quote}
\textit{``The system excluded many sentences here, but it's still very relevant... I am thinking how I can revert it back. It is hard to notice all filler words and delete them.''}
\end{quote}
P1 remarked, \textit{``I cannot add words instead of other words,"} when attempting to replace a repeated term with a pronoun, noting that this limitation \textit{``makes it hard to retain content.''}
%P5 comments on the ability for inserting new words within \system{} that \textit{``the user is more satisfied because of the larger amount of control they have for the transcript editing.''}
P11 adds that they would also insert new voice by using speech generation services, however, highlighting that \textit{``fine-tuning it''} to fit into the existing speech is time-consuming.
P5 comments on the ability for inserting new words within \system{}:
\begin{quote}
\textit{``[Editing in \system] is more satisfying because of the larger amount of control when editing the transcript.''}
\end{quote}
Reading their final edited transcript within \system's \textit{Final View}, P1 reported on the edited transcript coherence, \textit{``there is this good flow.''}
P11 stated that \system{}'s shortening is a \textit{``sharp tightening''} where \textit{``you are not missing any content.''}

P2 described that their usual process consists of \textit{``very low-level editing''} and that they required time to get familiar with more high-level edits that \system{} provided, as they \textit{``thought in too fine-granular edits''}. P10 mentioned that \system{} would be more helpful for editing unscripted speech than well-scripted speech, such as a politician's speech.
In addition to editing audio, 5 Participants (P1, P5, P9, P10, P12) mentioned they want to use \system{} for consuming audio, for example, to \textit{``re-watch past lectures.''} (P12).

\subsubsection{Editing Workflows}
All participants followed a consistent workflow across both interfaces.
Participants first chose an initial global compression rate, then skimmed and verified system edits segments by segment, and finally manually edited the speech to adjust content based on the editing purpose.
6 participants used the segment-level compression for local adjustments based on the segment's content relevance to the new purpose.
In the baseline interface, participants manually skimmed the transcript word by word, often describing this as \textit{``tiring''} (P6).
In contrast, \system{} revealed new editing strategies in participants' workflows.
%
% Participants included a similar amount of words removed by the system back in for both \system{} ($\mu$ = 120.36, $\sigma$ = 158.7) and the baseline interface ($\mu$ = 131.36, $\sigma$ = 165.43).
%
\\

\noindent \textbf{Choosing an initial compression.}
All participants first decided on an initial global compression.
In the baseline interface, participants had to skim the transcript for each compression to assess its quality.
P6 expressed particular frustration, describing the compressions as difficult to review: \textit{“I don’t really know what it does”}.

In \system{}, participants used the edit types view to quickly skim and review system edits.
For instance, P3 noted that the edit types view allowed them to quickly \textit{``get an overview''} of the edits and that it helped to choose an initial compression.
P4 found the edit types view particularly useful to find a balance between compression and preserving content: \textit{``A 30\% to 50\% compression seems like a good balance [...] to keep the information removal very low''}.
P3 mentioned understanding the speech first before editing took \textit{``less effort''} when using \system's outline pane compared to the baseline interface.

Participants employed different strategies when choosing an initial compression.
Participants either began with a high compression so that they could focus on reviewing edits and re-including critical content that the system removed or participants started with a low compression to focus on finding further content that can be removed according to their editing purpose.
\\

\noindent \textbf{Verifying system edits.}
After choosing an initial compression, participants verified whether the system edits aligned with their editing goals.
P11 mentioned that edit verification is part of their typical workflow and that they usually in Descript to later decide whether to remove it or not.
Participants typically explored and verified edits linearly by going through the speech transcript from start to end, edit by edit.
However, the strategies for this process differed between the two interfaces.

In the baseline interface, participants skimmed the entire transcript word by word to verify edits.
Participants reported  \textit{``jumping back up and down a lot [between words] intense''} (P6).
In contrast, when using \system{}, participants prioritize edits based on perceived risk and importance to manage cognitive load when verifying system edits.
Participants navigated paragraph by paragraph using the outline pane's higher-level content information, then focused on specific edits using the edit types view.
P10 describes their strategy when using \system's outline pane:
\begin{quote}
\textit{
You can look at it [the outline], compress, and check if the sentence that got compressed is related to the main topic of the paragraph or not. And when you think it's okay, then you can listen through it, and [...] make sure it all sounds correct [...] like sculpting something.
}
\end{quote}

P6 compared \system{} to the baseline interface and highlighted that the color-coded edit types reduced their mental effort, as they could \textit{``quickly filter''} low-risk edits (green) while focusing their attention on more critical edits (red) such as information removal.
%Similarly, P4 identified the edit types view as the \textit{``most helpful part''}, when verifying edits because they \textit{``only need to focus on information removal.''}
% For instance, P4 described using the outline pane to locate less-important sections for further compression, commenting that it was \textit{“helpful to divide segments into sentences,”} which made the review process more manageable.
While some participants, like P9, initially found the color-coding \textit{``distracting,''} they eventually appreciated it for verifying edits.
P4 the edit types view the \textit{``most helpful part:''}
\begin{quote}
    \textit{It reduced my effort [...] I only need to focus on information removal and for the filler word removal I just need to take a glance and skim it.\\}
\end{quote}

\noindent \textbf{Performing manual edits.}
All participants performed manual edits in both interfaces to ensure that the final edited speech aligned with their editing purpose.

Participants described manual editing in the baseline interface as \textit{``tiring''} (P6) because of the need to go linearly through the whole transcript to locate and remove speech disfluencies or unnecessary content.
For instance, P4 noted that \textit{``[they] went through everything''} and \textit{``manually did all the work''} when using the baseline interface.

In contrast, participants used \system{} to perform edits at multiple levels.
Participants used the edit type and diff views to perform edits on the word level, the outline pane to perform content-based edits, and the compression sliders for segment-level edits.
Participants appreciated the segment level compression sliders because they allowed them to compress whole segments based on their content relevance.
% For instance, P4 used the outline pane to identify less-important sections of the transcript, describing their strategy as identifying \textit{``lighter blue''} areas (indicating lower importance) and deciding whether to compress or remove them.
P6 noted that the outline pane \textit{``strategically breaks long sentences down,''} into \textit{``thought threads.''}
Participants used the outline pane to identify less-important sections of the transcript that can be edited.
%P4 further compared their strategy to their experience using the baseline interface and describe how they used the outline pane in \system{} to \textit{``remove the non-important parts first instead of going through the paragraphs linearly.''}.
P4 compared their strategy using the outline pane to their strategy within the baseline interface:
\begin{quote}
\textit{``I look at parts with lighter blue and remove the non-important parts first [in ~\system] instead of going through the paragraphs linearly [in the baseline].''}
\end{quote}
Participants highlighted the alignment between \system's importance indicators and their own judgment.
P3 felt that the system’s suggestions helped validate their manual edits: \textit{``I find it cool that the importance agrees with me when deleting.''}
However, while some participants, like P4 and P7, actively used the importance toggle to find additional content to cut by rendering less important content in gray, others engaged with this feature less frequently (Figure~\ref{fig:event_visualizations}).

\subsubsection{User Concerns}
All participants preferred \system{} over the baseline interface. However, some participants had difficulties transitioning between \system's views (P3, P9, P12).
For instance, P3 mentioned that it was \textit{ ``difficult to switch''} between views when losing their place in the transcript after switching from the diff view to the final view. We have since improved the interface to maintain cursor and scroll position when shifting views.
% , which suggests future interface improvements to improve user experience\amy{right now it is so vague that it is not worth saying... this clause could be applied to almost any limitation... if you're going to say anything, something like "final view. We have since improved the interface to maintain cursor and viewport position when changing views"}.
Although participants appreciated having more editing control in \system, participants raised concerns about being able to change a speaker's speech by making them \textit{``say something they never said.''} (P10).
P5 and P10 highlighted the ethical implications of editing speech to potentially misrepresent the speaker's intent or meaning, such as removing words of the professor's speech to \textit{``make it feel like [the professor] didn't care for [their] class''} (P10).
Although \system’s outline pane is aligned with the content in the compression pane to make editors aware of modifications in narratives, future iterations can flag edits that substantially change the original text's meaning.
% from the original speaker’s intent or risk propagating misinformation.

\section{Exploratory Study}
\revision{
The comparison study demonstrated that editors preferred \system{} over an existing extractive audio summarization baseline~\cite{wang2022record} for creating concise speech recordings.
Prior work indicates that synthesized speech can be indistinguishable to listeners from human speech for unfamiliar speakers~\cite{rosi2025, peng2024voicecraft, lu2025we}.
Thus, creators could theoretically use abstraction only for rewriting, then synthesize their entire audio without impacting audience perception.
However, it remains unclear whether creators themselves are comfortable with such extensive synthesis of their own speech.
As a result, we compare \system{} with the strongest abstractive audio summarization baseline we identified (GPT-4~\cite{openaiapi} + ElevenLabs~\cite{ElevenLabs}). We conducted a 60 minute exploratory study with three creators C1–C3\footnote{We recruited audio creators from our professional networks. One creator was male, two were female, and creators' ages ranged from 21–23.} who had more than 1 year of audio editing experience (for social media videos, podcasts, or class projects).
}
Creators first recorded a 2-3 minute speech recording\footnote{Creators selected a topic from a list of ten (see~\ref{sec:exploratory-prompts})}, then edited their speech using \system{}. We provided \system{} result and a fully synthesized version of the same result using state-of-the-art voice cloning~\cite{ElevenLabs}.
% We provided creators with their edited result produced both with \system{}'s algorithm and with a fully synthesized version generated through state of the art voice cloning (ElevenLabs~\cite{ElevenLabs}), then conducted a semi-structured interview. 
% The study took 60 minutes. 
% All creators had at least 1 year of prior experience editing their own speech for social media videos, podcasts, or class projects.
% Before the study, creators selected one topic from a provided list of 10 topics and recorded a 2-3 minute speech recording.
% Each creator then participated in a 60-minute, in-person session where they first answered background questions, received a tutorial on \system{}, and edited their speech using \system{}.
% We then showed creators the result produced by \system{} alongside a fully synthesized version generated through voice cloning by ElevenLabs~\cite{ElevenLabs}, followed by a semi-structured interview.

When comparing \system's result with the fully synthesized speech for the same transcript, all creators strongly favored \system's result as it preserved most of their authentic speech characteristics.
Creators described fully synthesized results as robotic (C1, C2, C3), monotone (C1, C2), and lacking personality (C2).  C1 noted \textit{``the way I talk is something I want to include,''} and mentioned that fully synthesized speech mispronounced uncommon terms in their speech, such as names of climbing walls. C3 said that for synthesized speech \textit{``I know it's not me''}, and they preferred \system{} because it preserved a \textit{``sense of authenticity''} as \textit{``the idea is to reuse and not regenerate''} speech.
While creators preferred \system's result and found most of \system's edits to be seamless, all creators noted at least one audible cut. 
All creators liked the automatic edits suggested by \system{} and made only minor edits (all inserted and removed at least one word). C1 expressed a desire for additional controls, such as the ability to manually insert pauses, and C3 described that their openness to synthesized edits depended on the context --- synthesis was appropriate for casual scenarios and audience who did not know their voice (\textit{e.g.}, posting to TikTok) but not for integrity-sensitive scenarios (\textit{e.g.}, formal class assignments). 
% Creators emphasized the importance of retaining their original vocal expressions, tone, and speech patterns. For example, C1 remarked, \textit{``the way I talk is something I want to include,''} and noted that fully synthesized speech mispronounced uncommon terms in their speech, such as names of climbing walls.
% halls\amy{walls?} they were mentioning in their original speech recording.
% C3 summarized their preference for \system{} by emphasizing authenticity:
%  \begin{quote}
%  \textit{``I think there has to be a sense to authenticity to it [...] the idea is to reuse and not re-generate [...] here [in \system] I still have my assets in.''}
%  \end{quote}
% All creators liked the automatic edits initially generated by \system{} as they sounded seamless, but all creators  and only made minor refinements.
% % to the system edits.
% Creators appreciated \system's seamless edits and mentioned hardly noticing audio cuts or synthesized insertions.
% However, all creators noted occasional abrupt transitions.
% C1 expressed a desire for additional controls, such as the ability to manually insert pauses.
% All creators chose to selectively synthesize minor connecting words or grammatical corrections, provided the major part remained original speech.
% C3 described contextual nuances, indicating comfort with synthesized speech in informal scenarios or when listeners are unfamiliar with their voice (\eg marketing videos on TikTok), yet strongly prefer to preserve their original speech for authenticity in integrity-sensitive contexts (\eg formal presentations and class assignments).
Overall, all creators wanted to use \system{} in the future to edit their speech recordings to balance authenticity with conciseness.
% \amy{ideally we could cut down this paragraph more }

%%%% sections/04-evaluation.tex ends here %%%%

%%%% 05-discussion.tex starts here %%%%

\section{Discussion}

\subsection{Reflection on Results}
Our evaluations demonstrate that \system{}'s algorithm and interface effectively address our design goals. \\ 
%by efficiently surfacing and removing unnecessary speech (\textbf{G1}), preserving key information (\textbf{G2}) and speaker style (\textbf{G3}), and providing granular control over automated edits that significantly reduce editing effort (\textbf{G4}), all while minimizing audio errors (\textbf{G5}).\amy{this seems to just restate our design goals}.

\noindent \textbf{TalkLess Algorithmic Results: }
Compared an extractive baseline~\cite{wang2022record}, \system{} produces shortened speech that preserves significantly more content for 15\%, 25\%, and 75\% compression targets and removes significantly more speech disfluencies while having significantly fewer coherence errors (\textbf{G1} -- surface and remove unnecessary speech, \textbf{G2} -- preserve important info, \textbf{G5} -- avoid audio errors).
\system{} preserves transcript style similarly as well as our extractive baseline and significantly better than a fully abstractive baseline (\textbf{G3} -- preserve speaker style). 
Human evaluators preferred \system{}'s outputs significantly more than baseline outputs for target compressions of 15\% and 25\%, while preferences were similar for higher compression rates that demanded more cuts and insertions. Creators in our exploratory study unanimously preferred editing their own speech with \system{} over complete re-synthesis, favored re-using original speech for authenticity (\textbf{G3}). \\
\noindent \textbf{TalkLess Interface Results: }All participants in our user evaluation preferred \system{} over the baseline text-based editor with extractive summary results.
With the baseline, participants edited by manually skimming through the transcript word by word, which was perceived as tedious and time-consuming. 
Participants thus rated \system{} as significantly easier and faster to use, less mentally demanding, and more engaging than the baseline due to \system{}'s visual skimming and reviewing aids that aligned with participants’ editing strategies (\textbf{G1}, \textbf{G2}, \textbf{G4}).
The edit types view, importance indicators, and outline pane specifically allowed participants to quickly identify and prioritize edits such that they could focus on meaningful content modifications rather than low-level manual edits.
\system{} thus significantly reduced manual edits \system{} (average of 478 manually edited words in \system{} compared to 807 words in the baseline) for a similar level of compression. \\ 
% \amy{is this in the results, and if so where?}\karim{-> end of editing workflows -- have to find the table with these numbers again to sanity check...}.
% Participants found that \system{} improved their editing workflow by allowing them to focus on significant content modifications rather than trivial edits.

\noindent \textbf{Use Considerations: }
Our evaluations also revealed ethical concerns that \system{} has the potential to unintentionally alter a speaker’s intended meaning when inserting or rearranging content, and context-specific acceptance or rejection of voice synthesis. \enlargethispage*{12pt}

\subsection{Limitations}
% While our system shows promise in improving speech transcript editing workflows, it is not without limitations.
\revision{We used ROPE~\cite{wang2022record} for sentence-level extraction in our baseline interface (Figure~\ref{fig:baseline_interface}).
While a well-engineered LLM-based extractive approach could surpass our baseline, developing such a system was beyond the scope of this work.}
Although VoiceCraft~\cite{peng2024voicecraft} represents the state-of-the-art in speech editing, it operates at low fidelity (16kHz sample rate) and can have unstable performance, such as dropped or distorted words during generation, which may negatively affect audio quality at higher compression rates (\textbf{G5}).
% These issues negatively affected audio quality in our evaluations, especially at higher compression rates and thus impacted listening preference.
Future speech editing models with higher fidelity and stable generation would increase the quality of results produced by \system.
% Integrating a higher-fidelity speech editing model (\eg models supporting at least 44.1 kHz sample rates) and improved generation stability would increase the quality and user acceptance of results produced by \system.
\system{} segments the original transcript into segments and compresses segments individually.
% However, this may overlook the global context of the transcript and important cross-segment relationships might be missed, potentially leading to a loss of content relationships in the edited audio.
To prevent missing global context and important cross-segment content relationships, we plan to explore hierarchical summarization techniques~\cite{Li2021hierarchicalspeechsummarization} in the future.
% To address this limitation, we plan to explore hierarchical summarization techniques~\cite{Li2021hierarchicalspeechsummarization}, enabling better integration of both local and global context in future iterations.
% Empirical experiments revealed that the number of unique edits plateaus after generating 25 candidates.
Finally, more optimal transcript edits may be identified by combining edits across edit candidates or generating more edit candidates and we will expand our candidate generation approach in the future. 
% \system's shortening pipeline also misses opportunities to combine edits from multiple candidates to increase the number of edit combinations to consider in the optimization algorithm.

% Although human evaluators appreciated the quality of \system's outputs, evaluators expressed comments on laughs or environmental sounds that can be automatically removed from the audio in future iterations by detecting noise, coughs, laughs, or breaths~\cite{dumpala2017}. 

% and improve the overall quality of the shortened speech transcript.

% Although our approach does not consider background noises, such as coughing or environmental sounds, human evaluators appreciated the overall quality of \system's outputs.
% However, human evaluators expressed comments about laughs in the background or noise from other distant persons.
% Hence, future integration of noise reduction and anomaly detection algorithms could improve the robustness and perceived professionalism of edited audio, especially in noisy or low-quality recording environments.
% Incorporating noise reduction and anomaly detection algorithms could enhance the system's robustness and produce cleaner audio results, for instance, through automatic removal of a speaker's laugh or a dog barking in the background.
% Allowing editors to make speech-rate adjustments (\eg speeding up or slowing down speech without pitch distortion) could provide greater flexibility when shortening speech recordings.
% Neural time-stretching algorithms~\cite{morrison2021neural} have shown promise to achieve these adjustments naturally.

\subsection{Extensions}
We demonstrate two extensions of \system{} that preview application scenarios and future directions.
% : moving along the spectrum of extractive-abstractive speech editing, and editing to preserve emphasis.
\\

\begin{figure}[h]
    \centering
    \includegraphics[width=\linewidth]{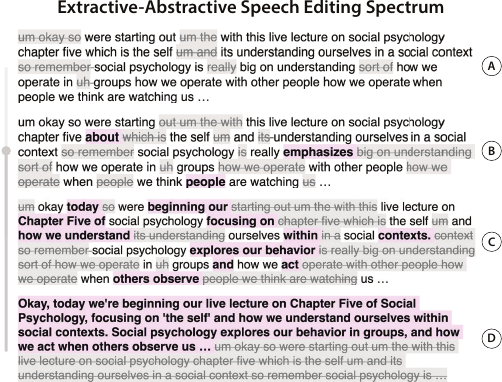}
    \caption{Extractive and abstractive speech editing creates a spectrum that allows partial blending between the two contrary editing types. Editors can choose between extractive (A), extractive-abstractive (B), abstractive-extractive (C), and abstractive (D) editing.}
    \Description{The image presents four versions of a speech transcript labeled A through D, illustrating a spectrum from extractive to abstractive editing. The heading reads "Extractive-Abstractive Speech Editing Spectrum." Version A, at the top, is mostly unedited and retains disfluencies such as filler words and repetitions. It includes phrases like “um okay so” and “uh,” and does not remove or paraphrase content. Version B introduces some minor edits. Certain disfluencies are struck through in gray, and some words are bolded, suggesting replacement or emphasis. The structure and wording remain close to the original, indicating a lightly edited extractive version. Version C shows a more heavily edited version. Many filler words and repetitions are removed or replaced, and more content is bolded and rewritten. It features a mix of original and reworded text, suggesting a midpoint between extractive and abstractive editing. Version D, at the bottom, is the most abstractive. It presents a polished, rewritten version in bold black text that summarizes the original content using new phrasing. The remaining portion of the original text is faded in gray and struck through, emphasizing the extent of rewriting. A vertical gray line with a circle at the top and bottom visually anchors the spectrum, guiding the viewer from minimally edited to fully rephrased content.}
    \label{fig:editing-spectrum}
\end{figure}

\noindent \textbf{Extractive-Abstractive Speech Editing Spectrum.}
Prior work explored the spectrum between extractive and abstractive methods for text~\cite{worledge2024extractive}.
Our work opens the opportunity to explore a similar spectrum for extractive and abstractive speech editing to balance preserving linguistic and para-linguistic speaker style with optimizing information compression, fluency, and clarity.
To support flexible editing strategies, we extended \system{} with a slider embedded directly within each paragraph in the compression pane (Figure~\ref{fig:editing-spectrum}).
Editors can select among four predefined editing modes:
\textit{Extractive} (Figure~\ref{fig:editing-spectrum}.A) to preserve maximum authenticity, \textit{extractive-abstractive} (Figure~\ref{fig:editing-spectrum}.B) to balance condensing information and preserving speaker style, \textit{abstractive-extractive} (Figure~\ref{fig:editing-spectrum}.C) that uses an abstractive summary for maximum content coverage but re-uses original speech to preserve speaker style \revision{(\eg Lotus~\cite{lotus2025} uses mostly abstraction to maximize compression but uses extraction when needed to preserve speakers' talking heads)}, and \textit{abstractive} (Figure~\ref{fig:editing-spectrum}.D) which synthesises the full abstractive summary. Each mode is implemented by explicitly instructing the LLM to do edits according to the chosen mode and adjusting the scoring algorithm to consider the balance between extractive and abstractive edits.

Future work could automatically recommend optimal slider settings based on speech characteristics (\eg spontaneous vs. scripted speech) and contextual constraints (\eg time limits), or user-specific editing styles learned from past interactions. Additionally, we plan to investigate when editors prefer voice re-synthesis versus re-recording or re-using audio to let our system automatically adjust abstractive-extractive sliders based on the editing context.
\\

\begin{comment}
\noindent \textbf{Content-based edits.}
\system{} allows editors to specify an editing purpose, which updates importance scores in the outline pane to surface less relevant or tangential content.
We extended \system's algorithmic methods to include the importance score based on the user's editing purpose in our automatic shortening pipeline.
Incorporating an additional importance parameter in the scoring function, \system{} can now choose edited candidate segments that remove only a little important but most of the irrelevant, tangential content.
We include additional candidate segments by prompting the underlying LLM to specifically remove major information chunks from the transcript based on their importance score.
Including these additional edited candidate segments in the selection pool allows \system{} to choose candidate segments with major content edits.\amy{do we have an example of before/after or any ideas of when this succeeds or fails? right now, I'm not learning much -- more like we could've done this but we didn't!}
\\
\end{comment}

\begin{figure}[h]
\centering
\begin{minipage}{\linewidth} % Adjust width here
\raggedright
\small % Optional: reduces text size a bit
\textbf{Original:}\\
So \emphasiss{remember}, a lot of these \emphasiss{concepts} with \emphasisl{the} \emphasism{self} are \emphasism{uh} \emphasiss{really} basic.\\[0.5em]

\textbf{Before considering emphasis:}\\
A lot of \emphasiss{concepts} with \emphasisl{the} \emphasism{self} are basic.\\[0.5em]

\textbf{After considering emphasis:}\\
\emphasiss{Remember}, these \emphasiss{concepts} with \emphasisl{the} \emphasism{self} are \emphasism{uh} \emphasiss{really} basic.
\end{minipage}
\caption{We extended \system{} to also preserve emphasized speech when shortening speech recordings.}
\Description{The image shows three lines of text labeled “Original,” “Before considering emphasis,” and “After considering emphasis.” Each line contains highlighted words in yellow. In the “Original” line, the text reads: “So remember, a lot of these concepts with the self are uh really basic.” The words “remember,” “concepts,” “self,” “uh,” and “really” are highlighted in yellow. In the “Before considering emphasis” line, the text reads: “A lot of concepts with the self are basic.” The words “concepts,” “the,” and “self” are highlighted in yellow. The words “remember,” “uh,” and “really” are not present in this line. In the “After considering emphasis” line, the text reads: “Remember, these concepts with the self are uh really basic.” The words “Remember,” “concepts,” “self,” “uh,” and “really” are highlighted in yellow. The structure and content of this line more closely resembles the “Original” than the second line.}
\label{fig:emphasis-edit}
\end{figure}

\noindent \textbf{Editing for Preserving Emphasis.}
Speakers use para-linguistic cues (\eg volume, tone, and pauses) to emphasize key points in their speech~\cite{chambershandbook,jm3, morrison2024crowdsourced, giles2012speech, erickson1978howard}.
To help speakers preserve these often intentional communicative elements, we extended \system{} with an emphasis-preserving editing module that automatically detects and prioritizes emphasized speech.
We used a neural speech prominence estimator~\cite{morrison2024crowdsourced} to assign each spoken word an emphasis value from 0 (no emphasis) to 1 (high emphasis), then modified the \system{} scoring function to penalize edits that would remove speech with high emphasis. We updated the interface to visualize emphasis with a yellow gradient to help editors surface emphasized words when reviewing edits~(Figure~\ref{fig:emphasis-edit}).
 While our extended approach preserves intended emphasis (\aptLtoX[graphic=no,type=html]{\colorbox[HTML]{fce45a}{\textit{``Remember''}}}{\textit{``\emphasiss{Remember}''}}), it also preserves speech errors (\aptLtoX[graphic=no,type=html]{\colorbox[HTML]{fce45a}{\textit{``uh really''}}}{\textit{``\emphasism{uh} \emphasiss{really}''}}) as speakers often unintentionally emphasize filler or transition words when they hesitate.
 % filler words, repetition, or tangential content that editors typically want to cut.
Future work can distinguish between intentional and unintentional paralinguistic patterns to preserve emphasis only where it is meaningful.
\\

\begin{comment}
\noindent While our prior shortening approach would shorten this to \textit{``A lot of \emphasiss{concepts} with \emphasisl{the} \emphasism{self} are basic.''}, our extension instead preserves emphasized phrases and produces \textit{`` \emphasiss{Remember}, these \emphasiss{concepts} with \emphasisl{the} \emphasism{self} are \emphasism{uh} \emphasiss{really} basic.''}
However, as speakers unintentionally also emphasize filler words, repetition, or tangential content, this also preserves speech errors (\textit{``\emphasism{uh} \emphasiss{really}''}) that editors typically want to cut.
Future work is needed to distinguish between intentional and unintentional paralinguistic patterns to preserve emphasis only where it is meaningful.
\end{comment}

\subsection{Future Work}

\noindent \textbf{Implications for Speech Editing.}
% when are our tradeoffs not good?
% in what cases
\system{} combines extractive speech editing~\cite{wang2022record, pavel2020rescribe} with synthesized abstraction using voice cloning~\cite{peng2024voicecraft, ElevenLabs}, enabling efficient shortening of raw, unscripted speech. However, this approach may be less effective for scripted speech, where disfluencies and tangents are less common, and high-level extraction alone may suffice.
That said, even scripted speech often includes paralinguistic speech errors, such as mispronunciations or unintended emphasis. To fix such errors, we suggest future work to explore tools that go beyond transcript-based editing to offer editors fine-grained control over paralinguistic features to give editors control over not just what is said, but how it is said.
Another direction is to explore how speech style varies based on audience context~\cite{giles2012speech, erickson1978howard}. Editors often adapt content to fit different audiences~\cite{benharrak2024writer, mieczkowski2021ai, hung2025simtube, choi2024proxona}, and a similar adaptation likely exists in speech. Future systems could support audience-aware speech editing to suggest edits depending on the listener.
\\

% \noindent \textbf{Risks of Rewriting Speech.}
% We recognize the potential risks that may arise when editing speech with \system, such as inserting unintended narratives or spreading misinformation \cite{hutiri2024not}
% For instance, P5 and P10 in our user evaluation raised concerns such as altering someone's attitude towards listeners when modifying original speech narratives.
% Although \system's outline pane is aligned with the content in the compression pane to make editors aware of modifications in narratives, future work should explore additional mechanisms for flagging edits that substantially diverge from the original speaker’s intent or risk propagating misinformation.
% \\

\noindent \textbf{Generalization to Other Media.}
\system's approaches hold promise for applications beyond speech, such as video and text editing.
Many creators repurpose long-form videos (\eg on YouTube) to produce short-form videos (\eg on TikTok) with a similar goal of condensing content as presented in prior work \cite{lotus2025, wang2024podreels}.
Prior work in video summarization \cite{gygli2014creating, song2015tvsum} compresses a long series of frames into short ones, either by removing single frames or whole video segments \cite{jin2017elasticplay, truong2007video}.
As these cuts are based only on the visuals, this leads to unnatural cuts in the audio track.
Future work can explore how audio and video compression approaches can be combined to address these issues.
\system{} extends transcript-based speech editing \cite{rubin2013content, Descript, AdobeLabsBlink}, which brings speech editing closer to writing.
This suggests opportunities for future work to integrate elements from prior work on writing assistants into work on transcript-based audio and video editing systems and vice versa.
For instance, \system's editing types could support efficient review of AI writing suggestions for shortening drafty text, particularly when combined with executable, verifiable edits~\cite{laban2024beyond}.
In addition, \system's outline pane with importance scores could assist writers in quickly identifying segments or sentences to cut in long text documents, similarly to how editors in our studies used the outline pane for segment- and sentence-level cuts.
We expect future systems to narrow the gap between transcript-based speech editing and writing systems.\\

\noindent \textbf{Editing for Perfection versus Authenticity.}
\revision{
\system{} makes a tradeoff between polishing the speech transcript through abstraction and preserving the original speaker style through extraction.
There exists a general trade-off that editors need to make between using editing tools to reach perfection (\eg auto speech enhancement~\cite{su2019perceptually, su2021hifi, abdulatif2024cmgan, serra2022universal}, automatic filler word removal~\cite{descriptUnderlord, podcastle}, condensing speech~\cite{wang2022record, lotus2025}) and avoiding editing for authenticity to preserve unique context and identity.
In professional contexts such as small business ads or lectures, users may prioritize polishing for perceived professionalism and clarity and thus desire the removal of disfluencies and filler words. Conversely, in personal or intimate contexts, users may want to preserve their authenticity by retaining personal filler words that reflect speaker personality~\cite{laserna2014like}.
Thus, to preserve more of the speaker's personality, future iterations of \system{} can avoid removing unique speech patterns by extending the scoring function to reward generations that preserve a higher count of unique speech patterns based on user-defined regular expressions where creators can define what words to preserve.
Similarly, future work can explore using LLMs to implicitly learn speaker-specific speech patterns based on the transcript to prevent the removal of those for authenticity.
% Similarly, creators may want to preserve non-speech related acoustic features (\eg background sound in a street interview). Future iterations of speech editing systems should explore integrating acoustic matching algorithms to seamlessly integrate newly generated or recorded speech to preserve authenticity~\cite{su2020}.
}
\\

% similar to GP-TSM, paragraph-wise is similar to beyond text generation, philippes paper on edit verification etc.

\noindent \textbf{Ethical Considerations.}
\revision{\system{} allows editors to efficiently shorten audio recordings.
However, it introduces ethical risks, particularly in high-stakes scenarios such as journalism, legal proceedings, and academic communication where accuracy and authenticity are important.
The ability to rewrite, condense, or rephrase speech raises concerns about editors being able to intentionally or unintentionally misrepresent a speaker's original intent or remove important contextual information.
Prior work on deepfakes has shown how synthetic audio can undermine trust in media~\cite{shoaib2023deepfakes}.
For example, when editing quoted speech material, even subtle changes can risk misrepresenting the credibility of the source.
To mitigate these risks, future systems could allow editors to ``lock'' specific transcript passages such as quotations or factual claims to prevent the system from modifying these passages to ensure preserving the verbatim speech.
Additionally, automated claim detection~\cite{konstantinovskiy2021} can be used to flag high-risk edits where factual statements are modified, while generating fact-check explanations~\cite{kazemi2021extractive} can support editors to verify correctness before publication.
To increase transparency when content is shared or published, content credentials~\cite{strickland2024election} describing the edit history would help collaborators or future editors track the speech's modification history and audio watermarking~\cite{chen2023wavmark} would help audiences identify modified segments.}

\section{Conclusion}
We introduced \system{}, a system that flexibly integrates extractive and abstractive summarization to shorten speech while preserving the original content and style.
Our approach combines a novel speech shortening pipeline, an automated audio editing mechanism, and a state-of-the-art open-source speech editing model.
Our evaluations revealed that \system{} significantly reduces speech disfluencies and incoherence errors while retaining more content compared to a previous automatic shortening method.
All participants in our user study (N=12) preferred editing speech with \system{} over a baseline interface, reporting significantly lower cognitive load and significantly higher creative flexibility.
In the exploratory study (N=3), we also explored how creators edit their own speech using \system.
We hope our work inspires future work to balance abstraction and extraction in content-based editing tools.

\begin{comment}
The system's interface supports creators in making edit decisions at multiple levels of detail, reducing the effort required to shorten speech recordings.
Our results evaluation demonstrated that \system{} effectively reduces speech disfluencies and incoherence errors while retaining more content compared to previous automatic shortening methods like ROPE.
The results evaluation showed that listeners preferred our shortened speech at lower compression rates.
In the user study, all participants preferred \system{} over the baseline interface, reporting significantly lower cognitive load and expressing enthusiasm for using the system in future audio editing tasks.
They highlighted how \system{} streamlined their workflow, allowed them to focus on significant edits, and provided flexible control over the editing process.
These findings suggest that \system{} holds significant potential for flexible audio shortening, particularly in contexts with spontaneous and unscripted speech such as podcasts, lectures, and interviews.
\system{} empowers creators to produce concise, coherent, and engaging audio content with reduced time and effort.
\end{comment}

%%%% sections/05-discussion.tex ends here %%%%

\bibliographystyle{ACM-Reference-Format}

%%%% bibliography starts here %%%%


%\bibliography{references}%% Commented by merge tool

\appendix

%%%% 06-appendix.tex starts here %%%%

\section{Appendix}

\subsection{Prompt Templates}
\subsubsection{Prompt Template for Generating Shortened Candidates}
\label{sec:appendix_prompt_candidate}
We distilled principles that describe requirements for ideal shortened speech transcripts based on our design goals and include them in our prompt:
\begin{itemize}
    \item All filler words should be removed (\textbf{G1}).
    \item All repetitions should be removed to avoid redundancies (\textbf{G1}).
    \item There should be no loss of information (\textbf{G2}).
    \item The original style should be preserved (\textbf{G3}).
    \item No new words should be added besides when merging sentences (\textbf{G3}).
    \item The original wording should not be changed (\textbf{G3}). 
    \item Unique words used for emphasis should be preserved (\textbf{G3}).
    \item Do not change any word spelling (\textbf{G3, G5}).
\end{itemize}

We share the prompt template in Python that we used below (some formatting added for the paper). The few-shot examples also followed this template and were inserted between the system description and the final instruction, as indicated below. This format is expected by the OpenAI API (Chat Completion). \textcolor{teal}{Python variables are color-coded in blue.}

{\scriptsize
\begin{lstlisting}
{
    "role": "system",
    "content": [
        {
            "type": "text",
            "text": f'You are a helpful writing assistant. Your task is to edit my transcript according to the following guidelines:
            1. **No New Words**: Do not add any new words that are not already in the transcript, besides when merging sentences.
            2. **Same Wording**: Do not change the original wording.
            3. **Include Everything**: Ensure that no piece of information from the original transcript is left out.
            4. **Remove Filler Words**: Eliminate all filler words like "um" and "uh."
            5. **Preserve Style**: Keep the original language style intact; don't change the tone or formal/informal nature of the language.
            6. **Remove Repetitions**: Delete any repeated information to avoid redundancies.
            7. **Unique Words**: Keep unique/rare words as they may aid in memory when listening to the transcript
            8. **No Hyphens or Word Corrections**: Do not combine words using hyphens or introduce hyphens and do not change a single character of a word to correct it if the word itself would be kept.
            9. **Target Length**: Ensure the final output is not more than {<@\textcolor{teal}{target\_length}@>} words.
            
            Your response will not be formatted and will only contain the shortened transcript.',
        }
    ],
},
...few-shot examples...
{
    "role": "user",
    "content": [
        {
            "type": "text",
            "text": f"{<@\textcolor{teal}{segment}@>}"
        }
    ]
}
\end{lstlisting}
}

\subsubsection{Prompt Template for Classifying Edit Types}
\label{sec:appendix_prompt_edit-types}
This is the prompt template we used with the same formatting as indicated in \ref{sec:appendix_prompt_candidate}.

{\scriptsize
\begin{lstlisting}
{
    "role": "system",
    "content": [
        {
            "type": "text",
            "text": "I will give you a sentence before and after an edit. Your task is to describe the edit that was made. Your response should not be formatted and only contain the type of the edit. Pick among the following types: ['Filler Word Removal', 'Repetition Removal', 'Clarification', 'Emphasis Removal', 'Information Removal'].
            
            Here are definitions of the edit types:
            **Filler Word Removal**: Removing unnecessary words or phrases that do not add meaning to the sentence, such as "um," "like," "you know," or "basically." These words are often used as verbal pauses and can be omitted without changing the overall content.
            Examples: ...
                        
            **Repetition Removal**: Removing repeated words, phrases, or ideas that add no new value to the content. This type of edit removes repeated words for clarity and conciseness.
            Examples: ...
            
            **Clarification**: This type of edits make an unclear or ambiguous phrase more understandable by providing additional context or rephrasing the phrase.
            Examples: ...
            
            **Emphasis Removal**: Removes words or phrases used to stress or exaggerate a point. These may include words like "really" and "very".
            Examples: ...
            
            **Information Removal**: Removing details, facts, or sections of the content that are either irrelevant to the main information or not relevant in a different context.
            Examples: ..."
        }
    ],
},
{
    "role": "user",
    "content": [
        {
            "type": "text",
            "text": f"
                Before: {<@\textcolor{teal}{before}@>}
                After: {<@\textcolor{teal}{after}@>}
                ",
        }
    ],
}
\end{lstlisting}
}

\subsubsection{Prompt Template for Importance Rating}
\label{sec:appendix_prompt_importances}
This is the prompt template we used with the same formatting as indicated in \ref{sec:appendix_prompt_candidate}.

{\scriptsize
\begin{lstlisting}
{
    "role": "system",
    "content": [
        {
            "text": 'Rate from 1-10, how crucial are the following information points to the main text content based on the following purpose:

            Purpose: {<@\textcolor{teal}{purpose}@>}
            
            For example, when an information is important contextually but not central to the main content of the text, it should be rated low.
            
            Format your output as JSON with a list of ratings from 1-10 for each information bit.
            
            Example output: {"importances": [X, Y, ...]}" where X, Y, ... are numbers between 1 and 10 for each respective information bit in the order I gave them to you.',
            "type": "text",
        }
    ],
},
... few-shot examples
{
    "role": "user",
    "content": [
        {
            "text": f"
                Information Points: {<@\textcolor{teal}{information\_points}@>}
                Text: {<@\textcolor{teal}{text}@>}
                "
            "type": "text",
        }
    ],
}
\end{lstlisting}
}

\subsubsection{Prompt Template for Content Extraction}
\label{sec:appendix_prompt_content-extraction}
This is the prompt template we used with the same formatting as indicated in \ref{sec:appendix_prompt_candidate}.

{\scriptsize
\begin{lstlisting}
{
    "role": "system",
    "content": [
        {
            "text": 'Extract a list of all atomic information bits that occur in a text I will give to you. An information bit is an atomic essiential piece of information that occurs in the text. Align each information bit with the respective phrase from the original text where this information bit is coming from. Additionally, give each information bit an importance score from 1 to 10 based on the information bits importance within the overall text. Your output response should be formatted as JSON in the following way:
            
            {
                "information_bits": [
                    {
                        "title": X,
                        "alignment": Y,
                        "importance": Z
                    },
                    {
                        "title": X,
                        "alignment": Y,
                        "importance": Z
                    },
                    ....,
                    {
                        "title": X,
                        "alignment": Y,
                        "importance": Z
                    }
                ]
            }
            
            where X is the information bit, Y is the phrase from the original text that aligns the most with the text, and Z is the information bit\'s importance (1-10) within the text.',
            "type": "text",
        }
    ],
},
{
    "role": "user",
    "content": [
        {
            "type": "text",
            "text": f"{<@\textcolor{teal}{text}@>}"
        }
    ]
}
\end{lstlisting}
}

\begin{table*}
\begin{tabular}{@{}lllllllllllll@{}}
\toprule
  &    & \multicolumn{3}{l}{Compression Target Deviation $\downarrow$} &  & \multicolumn{3}{l}{\% Words Synthesized $\downarrow$}       &  & \multicolumn{3}{l}{Coverage $\uparrow$}                       \\ \midrule
    Target    &    Method      & $\mu$         & $\sigma$            & range                   &  & $\mu$        & $\sigma$           & range                   &  & $\mu$         & $\sigma$           & range                    \\ \midrule
15\%    & LLM-only & 33.75\%          & 16.12\%          & 0.40\%-73.05\%          &  & 9.17\%          & 6.55\%          & 0.00\%-34.39\%          &  & 66.54\%          & 9.15\%          & 43.48\%-92.65\%          \\
        & Optimization & \textbf{7.53\%}  & \textbf{2.85\%}  & \textbf{0.09\%-32.38\%} &  & \textbf{1.15\%} & \textbf{0.54\%} & \textbf{0.00\%-8.12\%}  &  & \textbf{87.30\%} & \textbf{2.84\%} & \textbf{66.38\%-97.40\%} \\
25\%    & LLM-only & 32.70\%          & 16.25\%          & 0.23\%-59.96\%          &  & 8.11\%          & 6.42\%          & 0.00\%-32.38\%          &  & 66.36\%          & 7.66\%          & 37.00\%-90.69\%          \\
        & Optimization & \textbf{8.42\%}  & \textbf{3.74\%}  & \textbf{0.00\%-21.56\%} &  & \textbf{1.19\%} & \textbf{0.80\%} & \textbf{0.00\%-9.80\%}  &  & \textbf{84.41\%} & \textbf{3.59\%} & \textbf{65.32\%-96.43\%} \\
50\%    & LLM-only & 17.78\%          & 9.18\%           & 0.00\%-35.40\%          &  & 9.51\%          & 7.26\%          & 0.00\%-37.27\%          &  & 61.30\%          & 6.30\%          & 39.97\%-89.44\%          \\
        & Optimization & \textbf{3.69\%}  & \textbf{2.41\%}  & \textbf{0.00\%-18.14\%} &  & \textbf{2.58\%} & \textbf{2.05\%} & \textbf{0.00\%-13.17\%} &  & \textbf{75.20\%} & \textbf{3.33\%} & \textbf{62.45\%-88.40\%} \\ 
75\%    & LLM-only & 2.76\%           & 4.00\%           & 0.00\%-60.14\%          &  & 14.22\%         & 7.33\%          & 0.00\%-42.67\%          &  & 52.04\%          & 6.22\%          & 33.05\%-92.36\%          \\ 
        & Optimization & \textbf{1.49\%}  & \textbf{1.18\%}  & \textbf{0.00\%-6.53\%}  &  & \textbf{6.98\%} & \textbf{4.32\%} & \textbf{0.00\%-21.97\%} &  & \textbf{60.22\%} & \textbf{2.86\%} & \textbf{43.97\%-75.57\%} \\ \midrule
Overall & LLM-only & 21.75\%          & 5.14\%           & 0.00\%-73.05\%          &  & 10.25\%         & \textbf{0.41\%} & 0.00\%-42.67\%          &  & 61.56\%          & 1.20\%          & 33.05\%-92.65\%          \\
        & Optimization & \textbf{5.28\%}  & \textbf{0.92\%}  & \textbf{0.00\%-32.38\%} &  & \textbf{2.98\%} & 1.49\%          & \textbf{0.00\%-21.97\%} &  & \textbf{76.78\%} & \textbf{0.32\%} & \textbf{43.97\%-97.40\%} \\ \bottomrule
\end{tabular}
\caption{
\system's optimization approach generates shortened candidates with lower deviation from the target compression, lower \% of words synthesized and higher coverage compared to an ablation method where we prompted an LLM alone using our base prompt~\ref{sec:appendix_prompt_candidate}.
}
\label{tab:ablation_table}
\vspace*{-10pt}\end{table*}

\subsection{List of prompts used in the exploratory user study}
\label{sec:exploratory-prompts}
Creators chose one topic from a list of ten to record a 2-3 minute speech recording:
\begin{itemize}
    \item Describe your morning routine on a typical weekday.
    \item Talk about your favorite meal to cook or eat. Why do you love it?
    \item Tell a story about a time you overcame a challenge.
    \item What’s one hobby or activity you enjoy, and how did you get into it?
    \item Describe your ideal vacation - where would you go and what would you do?
    \item If you could have any superpower, what would it be and why?
    \item Talk about a piece of technology you use every day and how it affects your life.
    \item What’s something you wish you had learned earlier in life?
    \item Describe a place (real or fictional) that feels peaceful or inspiring to you.
    \item Share your thoughts on how the internet or social media has changed communication.
\end{itemize}

\begin{table}
\resizebox{\linewidth}{!}{%
\begin{tabular}{lllll} 
\toprule
\textbf{Measure} & \textbf{Baseline Interface} & \textbf{\system{}} & \textbf{$Z$} & \textbf{$p$} \\ 
\midrule
\textbf{NASA TLX} &  &  &  &  \\
Mental Demand & $\mu = 4.17$, $\sigma = 0.94$ & $\mu = 3.08$, $\sigma = 1.16$ & $2.0$ & \alphaval{<}{.05} \\
Temporal Demand & $\mu = 3.58$, $\sigma = 0.90$ & $\mu = 2.67$, $\sigma = 0.89$ & $2.0$ &  \alphaval{<}{.05} \\
Effort & $\mu = 3.83$, $\sigma = 1.03$ & $\mu = 3.08$, $\sigma = 0.79$ & $2.1$ &  \alphaval{<}{.05} \\
Frustration & $\mu = 2.50$, $\sigma = 1.51$ & $\mu = 1.58$, $\sigma = 0.67$ & $2.1$ &  \alphaval{<}{.05} \\
&  &  &  &  \\
\textbf{Creativity Support Index} &  &  &  &  \\
Exploration & $\mu = 1.75$, $\sigma = 0.75$ & $\mu = 4.42$, $\sigma = 0.51$ & $3.1$ &  \alphaval{<}{.01} \\
Engagement & $\mu = 2.75$, $\sigma = 1.22$ & $\mu = 4.33$, $\sigma = 0.89$ & $2.8$ &  \alphaval{<}{.01} \\
Worth the Effort & $\mu = 2.67$, $\sigma = 1.07$ & $\mu = 4.50$, $\sigma = 0.67$ &$ 2.7$ &  \alphaval{<}{.01} \\
Expressiveness & $\mu = 1.42$, $\sigma = 0.90$ & $\mu = 3.75$, $\sigma = 1.14$ & $3.0$ &  \alphaval{<}{.01} \\
&  &  &  &  \\
\textbf{System Usability Scale} &  &  &  &  \\
Frequent Use & $\mu = 2.08, \sigma = 0.79$ & $\mu = 4.5, \sigma = 0.67$ & $3.0$ & \alphaval{<}{.01} \\ 
Easy to use & $\mu = 3.33$, $\sigma = 0.98$ & $\mu = 4.42$, $\sigma = 0.67$ & $3.1$ &  \alphaval{<}{.05} \\
Quick to learn & $\mu = 4.00$, $\sigma = 0.74$ & $\mu = 4.58$, $\sigma = 0.51$ & $2.2$ &  \alphaval{<}{.05} \\
\bottomrule
\end{tabular}
}
\caption{Analysis of task load, creativity support, and usability ratings using paired Wilcoxon Sign Rank tests. We applied Greenhouse-Geisser correction when the equal-variances assumption is violated (Mauchly’s test \alphaval{<}{.05}).}
\label{tab:ratings_significances}
\end{table}

\begin{table}
\resizebox{\linewidth}{!}{%
\begin{tabular}{llll} 
\toprule
\textbf{Compression} & \textbf{\system} & \textbf{ROPE} & \textbf{Abstractive} \\ 
\midrule
15\% & $\mu = 0.92$, $\sigma = 0.06$ & $\mu = 0.95$, $\sigma = 0.06$ & $\mu = 0.58$, $\sigma = 0.07$ \\
25\% & $\mu = 0.9$, $\sigma = 0.07$ & $\mu = 0.96$, $\sigma = 0.02$ & $\mu = 0.44$, $\sigma = 0.13$ \\
50\% & $\mu = 0.91$, $\sigma = 0.05$ & $\mu = 0.93$, $\sigma = 0.04$ & $\mu = 0.65$, $\sigma = 0.07$ \\
75\% & $\mu = 0.86$, $\sigma = 0.05$ & $\mu = 0.88$, $\sigma = 0.01$ & $\mu = 0.52$, $\sigma = 0.07$ \\
\bottomrule
\end{tabular}
}
\caption{Mean and standard deviation of cosine style similarities between abstractive, \system{}, and ROPE transcripts, each compared to the original transcript.}
\label{tab:cosine_similarities_means}
\end{table}

\begin{figure}[h]
    \centering
    \includegraphics[width=2.5in]{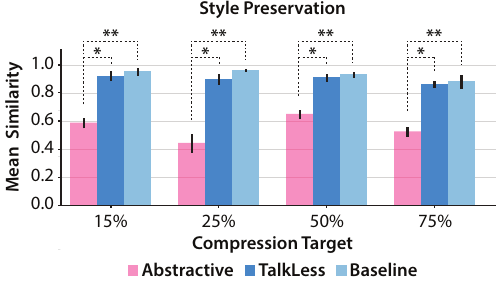}
    \caption{Cosine style similarities between abstractive, \system{}, and ROPE transcripts. Transcript styles of \system{} and ROPE were significantly closer to the original transcript style than the abstractive one for all compression rates but 25\%. *\alphaval{<}{.05}, **\alphaval{<}{.01}}
    \Description{A bar chart comparing mean cosine similarity scores for style preservation across three transcript methods—Abstractive, TalkLess, and Baseline—at four compression targets (15\%, 25\%, 50\%, and 75\%). TalkLess and Baseline consistently show higher similarity to the original transcript style than Abstractive, with statistically significant differences indicated at all compression targets except 25\%. Significance levels are marked with one star for a p value less than 0.05 and two stars for a p value less than 0.01}
    \label{fig:text-author-similarity}
\end{figure}

\begin{figure*}[h]
    \centering
    \includegraphics[width=0.85\textwidth]{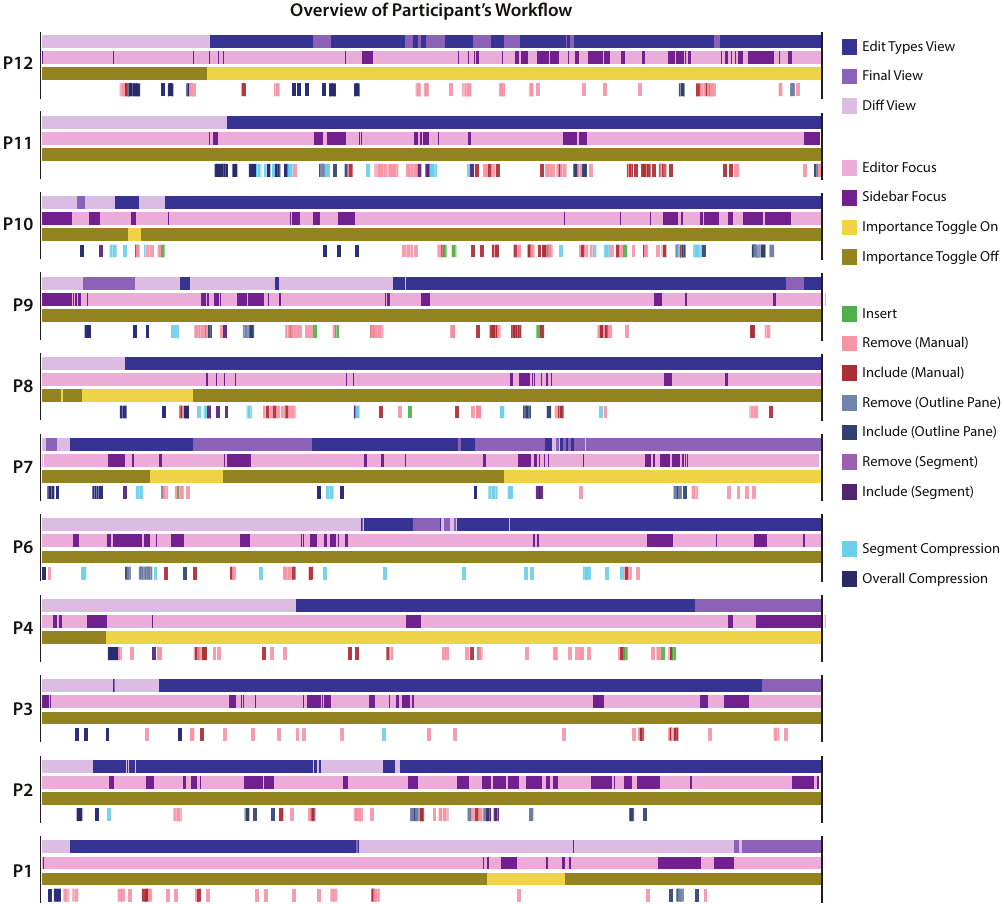}
    \caption{Interaction Behavior Overview.}
    \Description{A screenshot of the baseline interface used in the user evaluation study. The interface shows a transcript-based editor similar to TalkLess with a transcript of an American History class lecture. The transcription includes both spoken content and numerous edits. Strikethrough text represents parts of the original transcript that have been removed or modified during compression, while unstruck text remains in the final version. The top of the interface includes a compression slider, currently set at 20\%, with labeled stops at 0\%, 20\%, 47\%, 74\%, and a "Final" option. On the right side of the screen, there are two audio sections: “Original Audio” with a 20-minute timeline and audio controls, and “Generated Audio,” which is still loading. Two buttons beneath this section are labeled "Generate Audio" and "Log Transcript." The compressed transcript includes classroom greetings, logistical information, and historical context about the Cold War, the Gulf War, and U.S. military involvement, with whole sentences crossed out, indicating that they will be removed from the audio.}
    \label{fig:event_visualizations}
\vspace*{-10pt}
\end{figure*}

\begin{figure*}[h]
    \centering
    \includegraphics[width=7in]{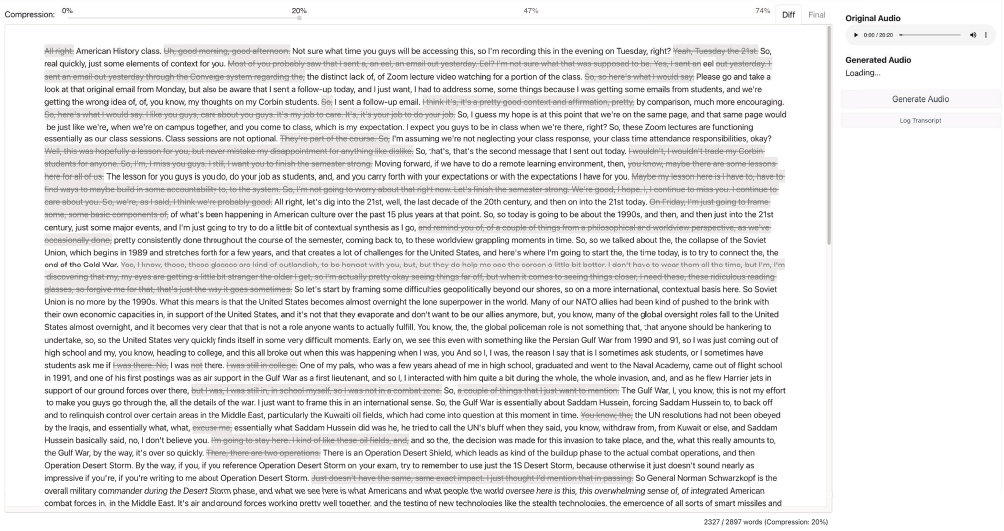}
    \caption{Baseline interface used in our user evaluation.}
    \Description{A detailed visualization illustrating the editing workflows of 12 participants (P1-P12) as horizontal timelines composed of colored segments. Each color represents different editing interactions, including views (Edit Types View, Final View, Diff View), editor and sidebar focus changes, toggling importance markers, insertion or removal of text (manual and segment-based), and actions related to segment or overall compression. Participants vary considerably in their editing patterns, with some frequently toggling views and importance markers, while others show distinct clusters of manual or segment edits.}
    \label{fig:baseline_interface}
\end{figure*}

\end{document}